\documentclass[preprint,12pt,nofootinbib]{revtex4-1}
\usepackage[paper=letterpaper,margin=1.5in]{geometry}

\usepackage{graphicx}\graphicspath{%
{}}%
\usepackage{amssymb,amsmath,amsfonts}
\usepackage{time}
\usepackage{epstopdf}
\usepackage{hyperref}
\hypersetup{
  colorlinks,
  citecolor=magenta,
  linkcolor=blue}
\usepackage[T2A,T1]{fontenc} % UTF-8: PDFLaTeX
\usepackage[utf8]{inputenc} % UTF-8: PDFLaTeX
\usepackage{txfonts} % UTF-8: PDFLaTeX
\usepackage[titletoc,title]{appendix}

\begin{document}

\begin{flushright}
{\normalsize
SLAC-PUB-16922\\
%arXiv submit/1732371\\
February 2017}
\end{flushright}

\title{Estimate of Joule Heating in a Flat Dechirper \footnote[1]{Work supported by the U.S. Department of Energy, Office of Science, Office of Basic Energy Sciences, under Contract No. DE-AC02-76SF00515
} }

\author{Karl Bane\footnote[2]{kbane@slac.stanford.edu} and Gennady Stupakov}
\affiliation{SLAC National Accelerator Laboratory,Menlo Park, CA 94025}
\author{Erion Gjonaj}
\affiliation{Technische Universit\"at Darmstadt, 64289 Darmstadt, Germany}

\begin{center}
%\today
\end{center}

\begin{abstract}

We have performed Joule power loss calculations for a flat dechirper. We have considered the configurations of the beam on-axis between the two plates---for chirp control---and for the beam especially close to one plate---for use as a fast kicker. Our calculations use a surface impedance approach, one that is valid when corrugation parameters are small compared to aperture (the perturbative parameter regime). In our model we ignore effects of field reflections at the sides of the dechirper plates, and thus expect the results to underestimate the Joule losses. The analytical results were also tested by numerical, time-domain simulations. We find that most of the wake power lost by the beam is radiated out to the sides of the plates. 
For the case of the beam passing by a single plate, we derive an analytical expression for the broad-band impedance, and---in Appendix B---numerically confirm recently developed, analytical formulas for the short-range wakes. 
While our theory can be applied to the LCLS-II dechirper with large gaps,
for the nominal apertures we are not in the perturbative regime and the reflection contribution to Joule losses is not negligible. 
With input from computer simulations, we estimate the Joule power loss (assuming bunch charge of $300$~pC, repetition rate of $100$~kHz) is 21~W/m for the case of two plates, and 24~W/m for the case of a single plate.

\end{abstract}

\maketitle

\section*{Introduction}

A corrugated structure device--a so-called dechirper~\cite{Bane12}--is being proposed for installation after the end of the linac and before the undulator regions of LCLS-II. Such a device has been installed in LCLS-I~\cite{Guetg16}, where it has been used for energy chirp control and as a fast kicker for self-seeding and two color operation of the FEL~\cite{Lutman16}. Because of the high repetition rate in LCLS-II compared to LCLS-I, Joule heating of the device by the beam's wakefields may now become significant and a cooling system may be required. Thus, it is important to estimate the amount of Joule heating power that is deposited in the corrugated plates.

%In previous work, the wake interaction between a short, high energy bunch and a flat dechirper was studied. For obtaining the Joule heating, however, since there are reflections at the edges of the dechirper plates, the longer time behavior of the fields in principle needs to be considered. However, since the frequencies of the dechirper are far above cut-off of the plates, one can make the approximation that such reflections are small and do not contribute significantly to the Joule losses. 

In previous work, the surface impedance approach to obtaining the impedances and wakes of a short, high energy bunch in a flat dechirper was derived~\cite{BaneStupakov12b,Zeta,Erratum}. The resulting formulas were shown to be approximate and valid in the perturbative regime, {\it i.e.} when $h/a\ll1$, with $h$ the corrugation depth and $a$ the half aperture of the dechirper. In this regime  (assuming also that $h/p\gtrsim1$, with $p$ the corrugation period) the impedance can be described by a single mode and the longitudinal wake by a damped cosine function~\cite{BaneStupakov03}. The RadiaBeam/SLAC dechirper at nominal parameters ($h=0.5$~mm and $a=0.7$~mm, thus $h/a=0.7$), however, is not in the perturbative regime. In this non-perturbative regime, it has been shown that the impedance consists of more than one mode~\cite{Zhang15} and the wake begins with a droop~\cite{Novokhatski15}. However, even for this regime analytical fitting functions for the short-range wakes have been derived and also verified by numerical simulation~\cite{BaneStupakovZagorodnov16}. 

%Nevertheless, for the nominal parameter regime, analytical fitting functions for the short-range wakes have been derived and also verified by numerical simulation~\cite{BaneStupakovZagorodnov16}. 

In previous work on the dechirper, the effect of the corrugations alone  was considered, and the effect of the resistance of the boundary metal was ignored. For Joule power loss calculations, however, one needs to include both contributions.
Also, in most of the previous work on the impedance of a flat dechirper, the case of the beam passing between two corrugated plates was considered. However, when used as a fast kicker, with the beam passing close to one plate, the second plate no longer influences the beam nor needs to be in the problem.  Note however, that analytical fitting formulas for the short-range wakes for this case have also been recently derived~\cite{one_plate_wakes}, though without numerical verification.

In the present report, by applying the surface impedance approach to both double and single plate dechirpers, we obtain analytical estimates of the Joule heating power. The surface impedance approach is used where it is not quite applicable ($h/a=0.7$ and is not small), introducing an error.  In addition, our estimates ignore the effect of reflections from the side edges of the dechirper plates. We thus expect our results to underestimate the Joule power losses. Finally, we test the accuracy of these calculations, by performing numerical simulations using the time-domain Maxwell equation solver in CST Particle Studio (PS)~\cite{CST}, and also (for verification purposes) with the program PBCI~\cite{PBCI}. These calculations are themselves quite challenging, since a fine mesh is needed to resolve the short bunch, and runs need to be performed over a large mesh domain for a relatively long time. In Appendix~B we perform numerical simulations to test the accuracy of the single plate, short-range wake formulas derived in~\cite{one_plate_wakes}. 

%, simulations that are challenging in themselves.%, requiring significant computational resources.
%Since the dechirper is envisioned to be used in two modes of operation---with the beam on axis, for chirp control, and with the beam near one jaw, to induce a fast transverse kick---we perform, in this report, calculations for both configurations.  
%In addition, we will derive and use formulas that are more accurate than those used before. 

%   Furthermore, wake calculations for the dechirper up to now have considered the effect of the corrugations alone, and ignored the effect of the resistance of the boundary material. 
%In this note, we derive and use formulas that are more accurate than those used before and---in order to estimate the Joule losses in the walls---they include both the effect of the corrugations and of the resistance in the walls. 

\begin{figure}[htb]
\centering
\includegraphics[width=0.5\textwidth, trim=0mm 0mm 0mm 0mm, clip]{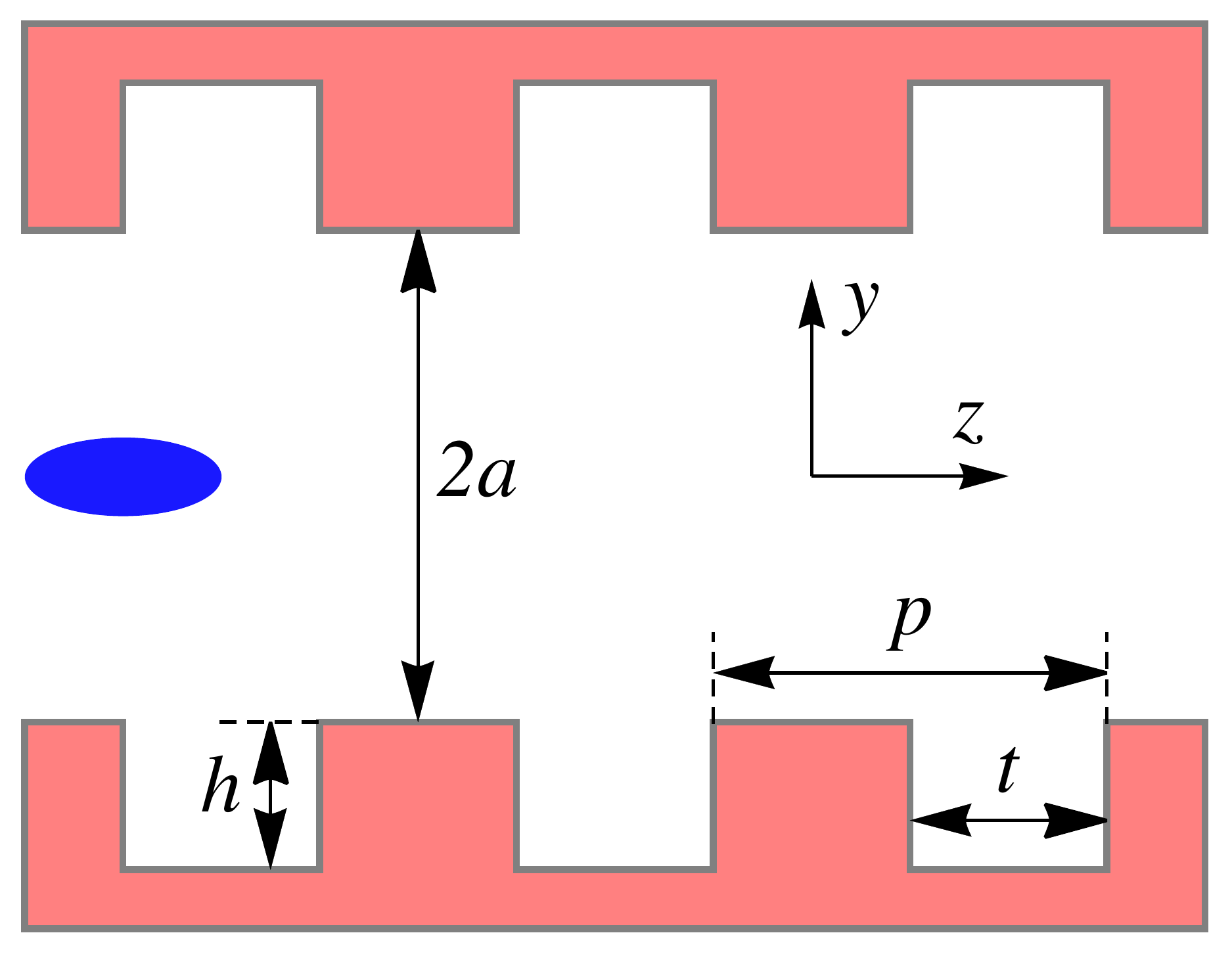}
\caption{Three corrugations of a vertical dechirper. A rectangular coordinate system is centered on the symmetry axis of the chamber. The blue ellipse represents an electron beam propagating along the $z$ axis.}
\label{fig:1}
\end{figure}

The RadiaBeam/SLAC dechirper that is installed in LCLS-I consists of two modules. Each module has two corrugated plates, with the beam passing between them (see Fig.~\ref{fig:1}). Two modules were chosen in order to partially cancel the unavoidable quadrupole wakefield that is induced by the beam; one has plates parallel to the $x$-$z$ axis (horizontal-longitudinal) plane and is adjustable vertically (the ``vertical dechirper"), and the complimentary one is adjustable horizontally (the ``horizontal dechirper"). For LCLS-II, the corrugation parameters are period $p=0.5$~mm, (longitudinal) gap $t=0.25$~mm, depth $h=0.5$~mm; nominally the gap $g=2a=1.4$~mm. The plate length $L=1.5$~m and width $w=12$~mm. Our calculations here consider: (i) a vertical two-plate dechirper, with the beam on the symmetry axis (such as is shown in Fig.~\ref{fig:1}); and (ii) only the top plate, with the beam just below it.
% (or, if there is a bottom plate, assuming that it is far away). 
The dechirper parameters and typical beam and machine properties used in calculations here are given in Table~I.

%Representative beam and machine parameters for the LCLS, which will be used in example calculations in this report, are given in Table~I. The wakes are given here in cgs units; to convert to MKS one merely needs to multiply by $(Z_0c/4\pi)$, with $Z_0=377$~$\Omega$ and $c$ is the speed of light. Wakefield induced energy loss, however, has been converted to MKS units, for convenience.

\begin{table}[hbt]
   \centering
   \caption{Selected beam and machine properties for LCLS-II used in example calculations. This is the high charge option with its maximum repetition rate. The charge distribution is assumed to be uniform with peak current $I=1.5$~kA. The dechirper properties are those of the RadiaBeam/LSLAC dechirper, which consists of two modules, each with corrugated plates of length $L=1.5$~m. The plates are made of aluminum; we take conductivity $\sigma_c=3.2\times10^{17}$/s.}
   \begin{tabular}{||l|c|c||}\hline\hline 
        {Parameter name} & {Value}  &  Unit\\ \hline\hline 
       Beam energy, $E$       &6  &GeV  \\
       Charge per bunch, $Q$       &300  &pC  \\
       Full bunch length, $\ell$       &60  &$\mu$m \\
       Repetition Rate, $f_{rep}$       &100  &kHz  \\
       Dechirper properties:       &   &  \\  
       Period, $p$ & 0.5 & mm \\
       Longitudinal gap, $t$ & 0.25 & mm\\
       Full depth, $h$ & 0.5 & mm\\    
       Nominal half aperture, $a$       & 0.7  & mm \\   
       Plate width, $w$       &12  & mm \\      
      Plate length, $L$       &2  &m \\
         \hline \hline 
   \end{tabular}
   \label{table1_tab}
\end{table}

 The equations in this report are given in cgs units. To convert an impedance or wake into MKS units, one multiplies the cgs result by $Z_0c/(4\pi)$, with $Z_0=377$~$\Omega$.

\section*{Joule Heating Estimates}

In a round corrugated structure of a finite length (a dechirper), the wakefield energy loss experienced by a relativistic beam of charged particles is partly absorbed in the walls as Joule heating and partly generates a THz pulse that leaves the structure just behind the driving particle. In the flat geometry of dechirpers like those that have actually been built---like the RadiaBeam/LCLS dechirper---some of that energy can also escape through the aperture to the side. 
The energy per unit length lost by the beam to the wake is then given by the sum
\begin{equation}
u_w=u_h+(u_{rad})_z+(u_{rad})_x\ ,
\end{equation}
with $u_h$ the energy generating Joule heating in the metal walls, $(u_{rad})_z$ the energy in the THz pulse that leaves the end of the structure following the driving particle, and $(u_{rad})_x$ the energy radiating out the sides of the structure.
%The time-domain, finite difference (wakefield solving) program ECHO(2D) can be applied to 2D rectangular structures (no boundary variation in the horizontal direction) and can include the effects of wall resistance~\cite{ECHO2D}. However, at the moment it is limited to conducting side walls, not sides with open boundary conditions. 
%Thus, to perform an accurate calculation of the amount of Joule heating that is generated by a train of bunches, one would need to use a 3D time domain program that can also account for resistive wall losses.

A particle of charge $Q$ moves at the speed of light $c$ on the axis of a structure. The Joule energy loss into the walls per unit length is given by
\begin{equation}
u_h=\frac{1}{c}\int_{\cal B} {\bf S}({\bf r},z)\cdot d{\bf A}\ ,
\end{equation}
with $S$ the Poynting vector, $d{\bf A}$ the incremental surface area vector (into the wall is positive), and $\cal B$ represents the metallic boundary. The calculation is performed at time $t=0$ when the particle is at $z=0$, and the transverse coordinate is $\bf r$. The particle is assumed to be moving to the left; the fields are zero ahead of the particle (for $z<0$).

Let us begin by sketching how we would solve the case of a dechirper with round geometry. The walls are located at radius $r=a$, and the Poynting vector at the walls is given by $S=-(\frac{c}{4\pi})E_zH_\phi$, with $E_z(r,z)$ the longitudinal component of the electric field and $H_\phi(r,z)$ the azimuthal component of the magnetic field. The Joule energy loss into the walls becomes
\begin{equation}
u_h=\frac{a}{2}\int_{-\infty}^\infty dz\,E_z(a,z)H_\phi(a,z)\ .\label{u_round_eq}
\end{equation}
%For the special case of a resistive wall impedance alone, A. Chao in his book gives explicit expressions for $E_z(r,z)$ and $H_\phi(r,z)$~\cite{AChao}. One can use these expressions to numerically obtain $u$ according to the above equation. The result, considering the conductivity of aluminum and a beam radius of $a=0.7$~mm, is that the Joule loss into the walls appears to equal the energy loss of the particles to the wake, $u_w$; {\it i.e.} $u_h=u_w=2Q^2/a^2$.
%This result is obviously true, since the wake fields have nowhere else to go but into the walls. To obtain the power into the walls (per unit length), we take $P=u_hf_{rep}$, with $f_{rep}$ the bunch repetition rate.
%This calculation is performed using a frequency domain approach. 
The fields can be written in terms of their Fourier transforms; for example, for the electric field 
\begin{equation}
\tilde E_z(\omega)=\frac{1}{c}\int_0^\infty dz\, E_z(z)e^{i\omega z/c}\ ,\quad E_z(z)= \frac{1}{2\pi}\int_{-\infty}^\infty d\omega\,\tilde E_z(\omega)e^{-i\omega z/c}\ ,
\end{equation}
where $\omega$ is the frequency and a tilde indicates the Fourier transform of a field (and similarly for $H_\phi$).
Substituting into Eq.~\ref{u_round_eq} and changing the order of integrations we obtain
\begin{equation}
u_h= \frac{a}{8\pi^2}\int_{-\infty}^\infty d\omega\,\tilde E_z(\omega)\int_{-\infty}^\infty d\omega'\,\tilde H_\phi(\omega')\int_{-\infty}^\infty dz\,e^{-i(\omega+\omega')z/c}\ .
\end{equation}
The last integral on the right equals $2\pi c\delta(\omega+\omega')$. Thus, we obtain
\begin{equation}
u_h= \frac{ca}{4\pi}\int_{-\infty}^\infty d\omega\,\tilde E_z(\omega)\tilde H_\phi(-\omega)= \frac{ca}{4\pi}\int_{-\infty}^\infty d\omega\,\tilde E_z(\omega)\tilde H_\phi(\omega)^*\ ,\label{u_round2_eq}
\end{equation}
where in the last integral $^*$ indicates the complex conjugate of a function. To obtain the last integral we used the relation $\tilde H_\phi(-\omega)=\tilde H_\phi(\omega)^*$; such a relation holds for the fields in the frequency domain, since the same fields in the time domain must be real quantities.

On the metallic surface we have the relation
\begin{equation}
\tilde E_z(\omega)=\zeta_z(\omega)\tilde H_\phi(\omega)\ ,
\end{equation}
with $\zeta_z(\omega)$ the surface impedance of the structure walls, which includes both the contributions of the corrugations and the wall resistance (the subscript $z$ indicates that the surface currents move in the longitudinal direction). Substituting into Eq.~\ref{u_round2_eq}, and noting the symmetry of the integrand, we finally obtain
\begin{equation}
u_h= \frac{ca}{2\pi}\int_{0}^\infty d\omega\,Re[\zeta_z(\omega)]|\tilde H_\phi(a,\omega)|^2\ .\label{u_roundc_eq}
\end{equation}
If we write the Joule loss as $u_h=\int_{-\infty}^\infty\frac{d\tilde u}{d\omega}d\omega$, then we can define an effective Joule heating impedance $Z_h(\omega)$, which has a real part defined as
\begin{equation}Re(Z_h(\omega))=\frac{\pi}{Q^2}\frac{d\tilde u}{d\omega}(\omega)\ .\label{ReZh_eq}
\end{equation}
In this round example it turns out that all the wake losses end up in the walls, since they have nowhere else to go. It is easy to see that $Z_h(\omega)=Z(\omega)$, with $Z(\omega)$ the normally defined impedance of the corrugated structure, given by
$Z(\omega)=-{\tilde E_z}/Q$.
For a bunch of particles, the Joule heating energy is given by
\begin{equation}
u_{h\lambda}=\frac{1}{\pi}\int_{0}^\infty d\omega\,|\tilde I(\omega)|^2Re(Z_h(\omega))\ ,
\end{equation}
with $\tilde I(\omega)$ the Fourier transform of the current, $I(z)=Qc\lambda(z)$, and $\lambda(z)$ the longitudinal bunch distribution. However, in following energy and power calculations we will let ${|\tilde I}(\omega)|^2\rightarrow1$, since for the small bunch lengths considered and the frequency reach of the dechirper impedance, this approximation is good.
%---the Table~I example has rms bunch length, $z_{rms}=17$~$\mu$m--
%the beam spectrum reaches to much higher frequencies than the dechirper impedance, and the effect is small.

%%The advantage of the spectral form of the equation for $u_h$ (Eq.~\ref{u_roundc_eq}), compared to the temporal form (Eq.~\ref{u_round_eq}), is that its integrand can be written explicitly for the case of a pipe that is both resistive and corrugated, at least if the corrugations are small enough that the perturbation solution holds. %We will here not carry the round calculation any further.%; after all, even with both resistance and corrugations, all the power lost by the beam to the wakes ends up in the walls, since it has nowhere else to go.

The Joule power loss is simply given by $P=u_hf_{rep}$, with $f_{rep}$ the bunch repetition rate. However, we need to make one more adjustment. In the round case, since all the beam energy loss becomes Joule heating, we obtain $u_h=2Q^2/a^2$. However, our Joule energy loss calculations are  perturbative calculations. Since in our calculations the corrugation parameter ratio $(h/a)$ is not small we are not in the perturbative regime and there will be wake droop. To better estimate the Joule power loss we take
\begin{equation}
P=f_{rep}u_h\left[\frac{\varkappa(\sigma_z)}{\varkappa(0)}\right]\ ,
\end{equation}
with $\varkappa(\sigma_z)$ the loss factor of a Gaussian bunch of length $\sigma_z$. The point charge loss factor $\varkappa(0)=2/a^2$ (with $a$ the pipe radius) in the round case, and $\varkappa(0)={\pi^2}/(8a^2)$ (with $a$ the half aperture) in the flat case.  The loss factor $\varkappa(\sigma_z)$ includes the effect of the wake droop.

\section*{Flat Geometry}

\begin{figure}[htb]
\centering
\includegraphics[width=0.55\textwidth, trim=0mm 0mm 0mm 0mm, clip]{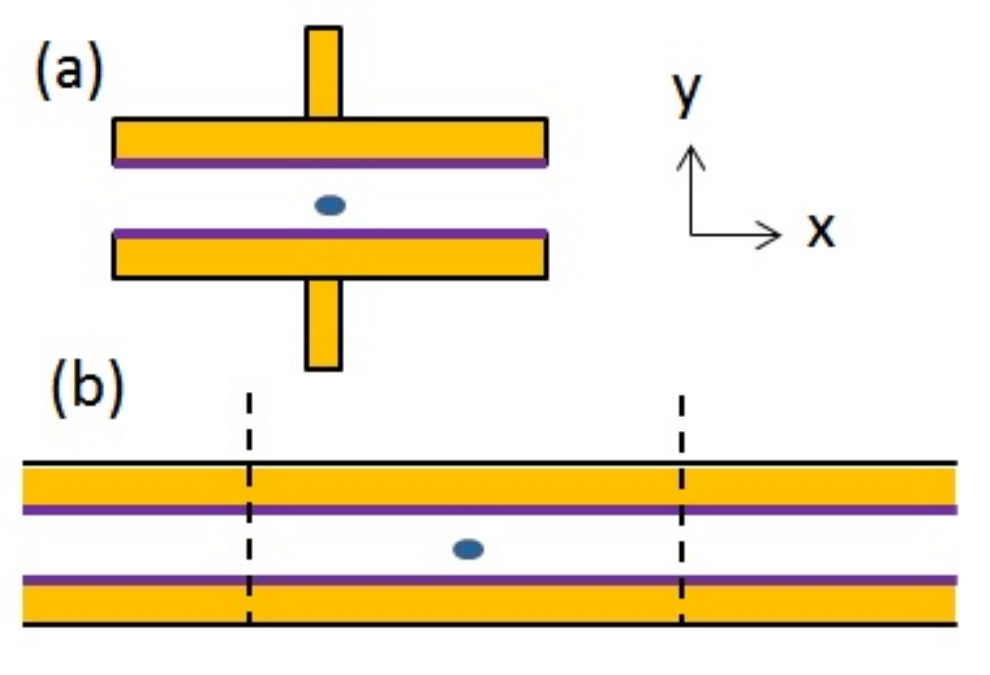}
\caption{Sketch of a vertical set of dechirper jaws seen end on: (a)~in the real geometry the jaws have a finite width $w$; (b)~in the model used in the calculation, the width is infinite, but the loss integration is performed only over a width $w$ (represented by the distance between the dashed lines). The purple lines represent the corrugated surfaces; the blue ellipses, the exciting particle's transverse location.}
\label{sketch_fi}
\end{figure}

In flat geometry energy can also radiate out the sides. To obtain our Joule heating estimate we perform a simple approximate calculation. End on, a vertical dechirper looks like what is shown in the sketch of Fig.~\ref{sketch_fi}a. The vertical gap is $2a$ and the width of the corrugated plates is $w$. For our calculations we let the width of the plates become infinite, and to account for the finite width as in the real case, we perform Joule loss calculations over the boundary only over a region of width $\pm w/2$ from the beam path (see Fig.~\ref{sketch_fi}b; the distance between the dashed lines is meant to be $w$). In addition, we use a perturbation calculation that is only accurate when the corrugation parameters are small compared to the aperture $2a$. For the parameters of Table~I we see that this is not true, so we will lose accuracy here, too. The final contributor to inaccuracy is that the perturbation calculation assumes that the structure is infinitely long, which does not allow for part of the wake loss to contribute to the generation of a THz pulse in the $z$ direction. However, this final contribution can be shown to be small for the parameters of Table~I.% in the Appendix. 

Let us consider a vertical dechirper with plate walls of width $w$ located at  $y=\pm a$ with respect to the axis. The Poynting vector at the walls 
\begin{equation}
S=\left(\frac{c}{4\pi}\right)(E_x H_z-E_zH_x)\ . 
\end{equation}
Note that in this case, in addition to a surface current in the $z$ direction, corresponding to the surface impedance $\zeta_z$, there is a surface impedance in the $x$ direction, corresponding to surface impedance $\zeta_x$.
The surface impedance in the $z$ direction is given by the sum of the resistive wall and the corrugation impedance contributions, $\zeta_z=\zeta_{rw}+\zeta_{corr}$, with~\cite{AChao}, \cite{BaneStupakov12b},
\begin{equation}
\zeta_{rw}(\omega)=\left(\frac{\omega }{8\pi\sigma_c}\right)^{1/2}(1-i)\ , \quad\quad\zeta_{corr}(\omega)=-i\frac{h\omega}{2c}\ ,
\end{equation}
and $\sigma_c$ is the conductivity of the metal walls. 
In the $x$ direction, we take the surface impedance to be $\zeta_x=\zeta_{rw}$, since the horizontal surface currents are not impeded by the corrugations. Note that an anisotropic surface impedance was not used before for modeling the impedance of the corrugated structure; this form, however, is important in our application here. 
Note also that with only corrugations and no wall resistance, the Joule heating is zero.

The Joule wall energy loss per unit length becomes
\begin{equation}
u_h=\frac{1}{2\pi}\int_{-w/2}^{w/2}dx\int_{-\infty}^\infty dz\,\left[E_z(x,a,z)H_x(x,a,z)+E_x(x,a,z)H_z(x,a,z)\right] \ 
\end{equation}
(an overall factor of 2 is added because there are two plates). There are two contributions. Let us take $u_h=u_{hz}+u_{hx}$, with $u_{hz}$ the part that depends on $E_zH_x$ at the walls; $u_{hx}$ the part that depends on $E_xH_z$ at the walls.
%Since the losses are the same on the top plate as on the bottom one, in the rightmost expression we integrate over only the top plate and multiply by a factor of 2. 
Following a calculation similar to that for the round case above, we can rewrite the equation for $u_z$ in the frequency domain as
\begin{equation}
u_{hz}= \frac{c}{4\pi^2}\int_{-w/2}^{w/2}dx\,\int_{-\infty}^\infty d\omega\,Re[\zeta(\omega)]|\tilde H_x(x,a,\omega)|^2\ .\label{u_flat_eq}
\end{equation}

To perform the calculation, we follow the procedure described in Ref.~\cite{Zeta}, which explicitly gives the fields and wakefields in structures with flat geometry for which the effect at the boundaries can be approximated by a surface impedance. First the frequency representation of the fields are Fourier transformed in $x$ as: 
\begin{equation}
\hat H_x(q,y,\omega)=\int_{-\infty}^\infty dx\,\tilde H_x(x,y,\omega)e^{iqx}\ ,\quad \tilde H_x(x,y,\omega)=\frac{1}{2\pi}\int_{-\infty}^\infty dq\, \hat H_x(q,y,\omega)e^{-iqx}\ .
\end{equation}
Substituting into Eq.~\ref{u_flat_eq}, changing the order of integration, and noting the symmetry of the integrand with respect to $\omega$, we find that
\begin{equation}
u_{hz}=\frac{cw}{8\pi^4}\int_{0}^\infty d\omega\,Re[\zeta_z(\omega)]\int_{-\infty}^\infty dq\,\hat H_x(q,a,\omega)\int_{-\infty}^\infty dq'\,\hat H_x^*(q',a,\omega)\,{\rm sinc}\left[\frac{w(q-q')}{2}\right]\ .\label{losses_eq}
\end{equation}
It is this triple integral that we solve numerically to estimate the fraction of wakefield losses that end up as Joule heating in the walls.
Note that for the special case width $w\rightarrow\infty$, the equation simplifies to
\begin{equation}
u_{hz}=\frac{c}{2\pi^3}\int_{0}^\infty d\omega\,Re[\zeta_z(\omega)]\int_{0}^\infty dq\,|\hat H_x(q,a,\omega)|^2\ ,\label{u1_winf_eq}
\end{equation}
(we have used the fact that the integrand is symmetric with respect to $q$).

Performing a similar calculation for $u_x$, we obtain, for finite $w$,
\begin{equation}
u_{hx}=\frac{cw}{8\pi^4}\int_{0}^\infty d\omega\,Re[\zeta_x(\omega)]\int_{-\infty}^\infty dq\,\hat H_z(q,a,\omega)\int_{-\infty}^\infty dq'\,\hat H_z^*(q',a,\omega)\,{\rm sinc}\left[\frac{w(q-q')}{2}\right]\ ,\label{u2_flat_eq}
\end{equation}
and for infinite $w$
\begin{equation}
u_{hx}=\frac{c}{2\pi^3}\int_{0}^\infty d\omega\,Re[\zeta_x(\omega)]\int_{0}^\infty dq\,|\hat H_z(q,a,\omega)|^2\ .\label{u2_winf_eq}
\end{equation}
%We denote here the surface impedance by $\zeta'$, to indicate that the surface impedance in the $x$ direction can be different than the one in the $z$ direction. 

The general form of the impedances for a flat structure, one that can be described using the surface impedance concept, is developed in Refs.~\cite{Zeta,Erratum}. A conclusion of this work was that the final results---the impedances---were approximate and valid, provided that  the frequency $k=\omega/c\gg 1/a$---which is satisfied---and that $|\zeta|\ll1$ (or $h/a\ll1$), which is not. The expressions for $\hat H_x(q,a,\omega)$, $\hat H_z(q,a,\omega)$, that we need here were not given in the earlier reports. Following the correct derivation of the fields, 
%in the Appendix of \cite{BaneStupakov16}, 
we obtained the general form of $\hat H_x(q,a,\omega)$ and $\hat H_z(q,a,\omega)$ as functions of beam offset (not shown).
% The case of beam location $y_0>0$ and test location $a\ge y>y_0$ is given in Appendix~xxx.
For the special case of the beam on axis, the fields on the boundary at $y=a$ are
\begin{eqnarray}
\hat H_x&=&-\frac{4\pi Q}{c}\frac{ i k q \cosh (a q)}{\zeta_z  (k-q) (k+q) \sinh (2 a q)+2 i
   k q \left(\zeta_z  \text{$\zeta_x $} \sinh ^2(a q)+\cosh ^2(a
   q)\right)}\ ,\nonumber\\
\hat H_z&=&-\frac{4\pi Q}{c}\frac{ \zeta_z  k q \sinh (a q)}{\zeta_z  (k-q) (k+q) \sinh (2 a
   q)+2 i k q \left(\zeta_z  \text{$\zeta_x $} \sinh ^2(a q)+\cosh
   ^2(a q)\right)}\ ,
\end{eqnarray}
%where $k=\omega/c$.
with $k=\omega/c$. %It is easy to see that with these fields inserted into Eqs.~\ref{losses_eq} and \ref{u2_flat_eq}, the symmetry with respect to $q$, $q'$, leads to real results for $u_1$ and $u_2$, as required.

%From the symmetries of the integrands in Eqs.~\ref{losses_eq} and \ref{u2_flat_eq} with respect to $q$, $q'$, we can easily see that these equations indeed give real results$.

\begin{figure}[htb]
\centering
\includegraphics[width=0.8\textwidth, trim=0mm 0mm 0mm 0mm, clip]{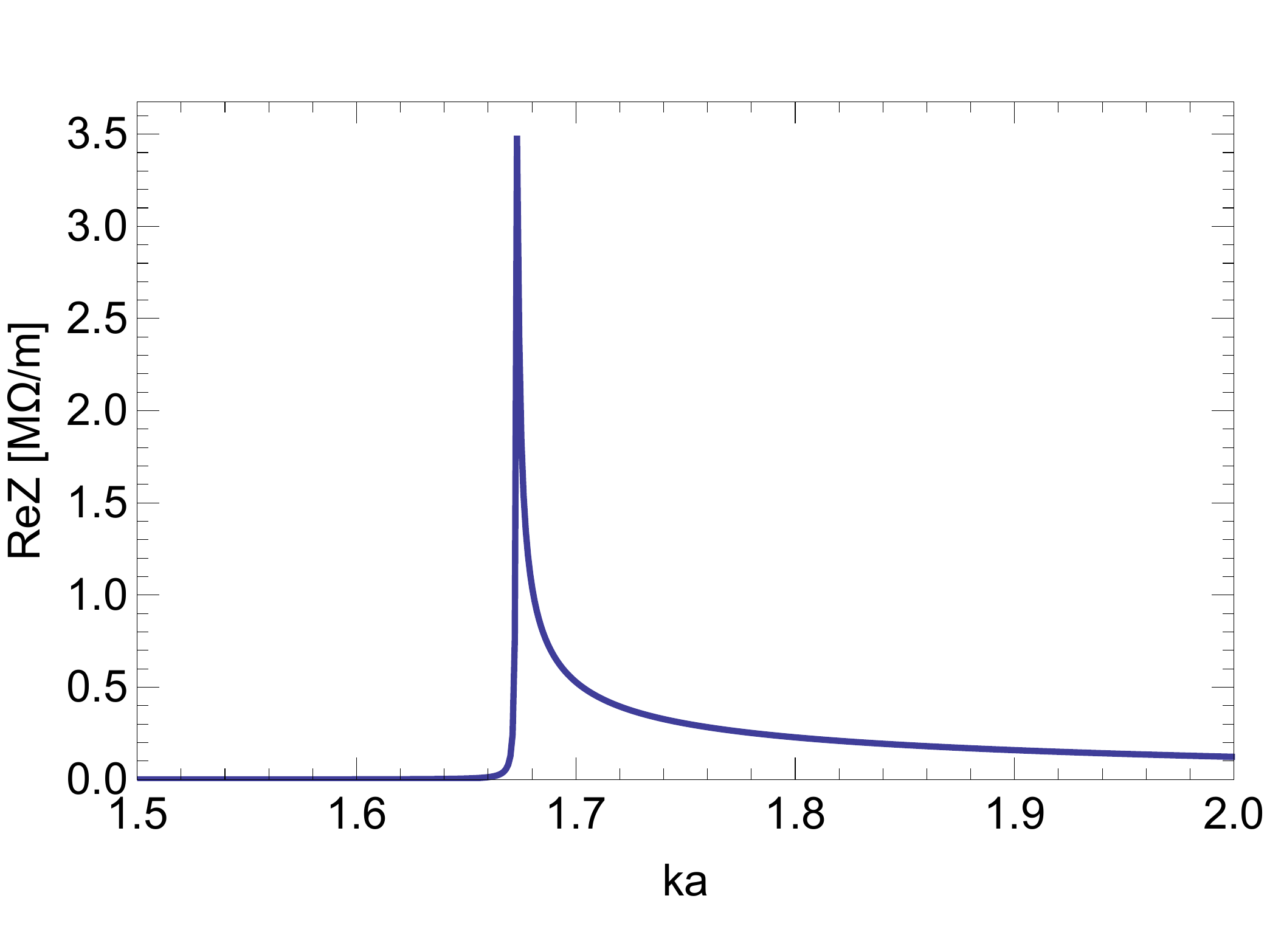}
\caption{The real part of the longitudinal impedance for the dechirper with plate width $w\rightarrow \infty$. This plot, in different units, and with the spike reaching to infinity, can also be found in Ref.~\cite{BaneStupakov03}.}
\label{Z_winf_fi}
\end{figure}

In the case of flat geometry, with corrugated plate width $w\rightarrow\infty$, the total Joule energy loss on the walls, $u_z+u_x$, must equal a point charge beam's energy loss on the axis, $u_w$. In this case the point charge loss factor
\begin{equation}
\kappa(0)=\frac{u_w}{Q^2}=-\frac{1}{\pi Q}\int_0^\infty d\omega\,Re[{\tilde E_z}(\omega)]=-\frac{1}{\pi^2 Q}\int_0^\infty d\omega\int_0^\infty dq\,Re[{\hat E_z}(\omega,q)]\ ,\label{kappa_eq}
\end{equation}
with the on-axis electric field
\begin{equation}
\hat E_z=-\frac{4\pi Q}{c}\frac{ i k q \zeta_z}{\zeta_z  (k-q) (k+q) \sinh (2 a q)+2 i
   k q \left(\zeta_z  \text{$\zeta_x $} \sinh ^2(a q)+\cosh ^2(a
   q)\right)}\ .\label{hatEz_eq}
\end{equation}
The impedance is given by
\begin{equation}
Z(\omega)=-\frac{\tilde E_z(\omega)}{Q}=-\frac{1}{2\pi Q}\int_{-\infty}^\infty dq\, \hat E_z(q,\omega)\ .
\end{equation}
Substituting from Eq.~\ref{hatEz_eq} and integrating numerically we obtain the impedance.
The real part, for this case ($a=0.7$~mm and $w\rightarrow \infty$), is shown in Fig.~\ref{Z_winf_fi}. We see that the impedance is a highly spiked function, with the peak location at $ka\approx\sqrt{2a/h}\approx1.673$~\cite{BaneStupakov03}. Note that a different, earlier perturbation analysis, one that ignored wall resistance, gave essentially the same result except that the spike reached to infinity~\cite{BaneStupakov03}. This implies that the short-range wake is essentially independent of the boundary conductivity.

%The function $Re(Z)$ is mostly determined by the corrugations. %In this calculation if one sets $\zeta_{rw}\rightarrow0$, $\hat E_z$ becomes purely imaginary (see Eq.~\ref{hatEz_eq}), and the integral over $q$ to obtain $\tilde E_z$ diverges. However, if the integral is taken as a Cauchy integral, 
%A perturbation calculation that ignores wall resistance was performed in Ref.~\cite{BaneStupakov03}; that calculation results in an almost identical function to that of Fig.~\ref{Z_winf_fi}, except that the peak now reaches to infinity (see figure in \cite{BaneStupakov03}).

 The Joule heating energy has two components, $u_{hz}$ and $u_{hx}$, corresponding to contributions from $\zeta_z$ and $\zeta_x$, respectively. Numerically solving Eqs.~\ref{u1_winf_eq} and \ref{u2_winf_eq} for the infinitely wide plates, we find that $u_{hz}=0.8u_w$,  $u_{hx}=0.2u_w$, and, to good accuracy, $u_h\equiv u_{hz}+u_{hx}=u_w=Q^2{\pi^2/(8a^2)}$. This is what we expect: for plates of infinite width, the sum of the two Joule energy contributions should equal the energy loss of the on-axis point charge beam. 
%From Eq.~\ref{ReZh_eq} we obtain the real part of the Joule heating impedances, and find that $Re(Z_{hz}(\omega))\approx Re(Z(\omega))$ and $Re(Z_{hx}(\omega))\approx0$. %We find that %Zooming in, however, we can see details of the two different contributions. 

\begin{figure}[htb]
\centering
\includegraphics[width=0.8\textwidth, trim=0mm 0mm 0mm 0mm, clip]{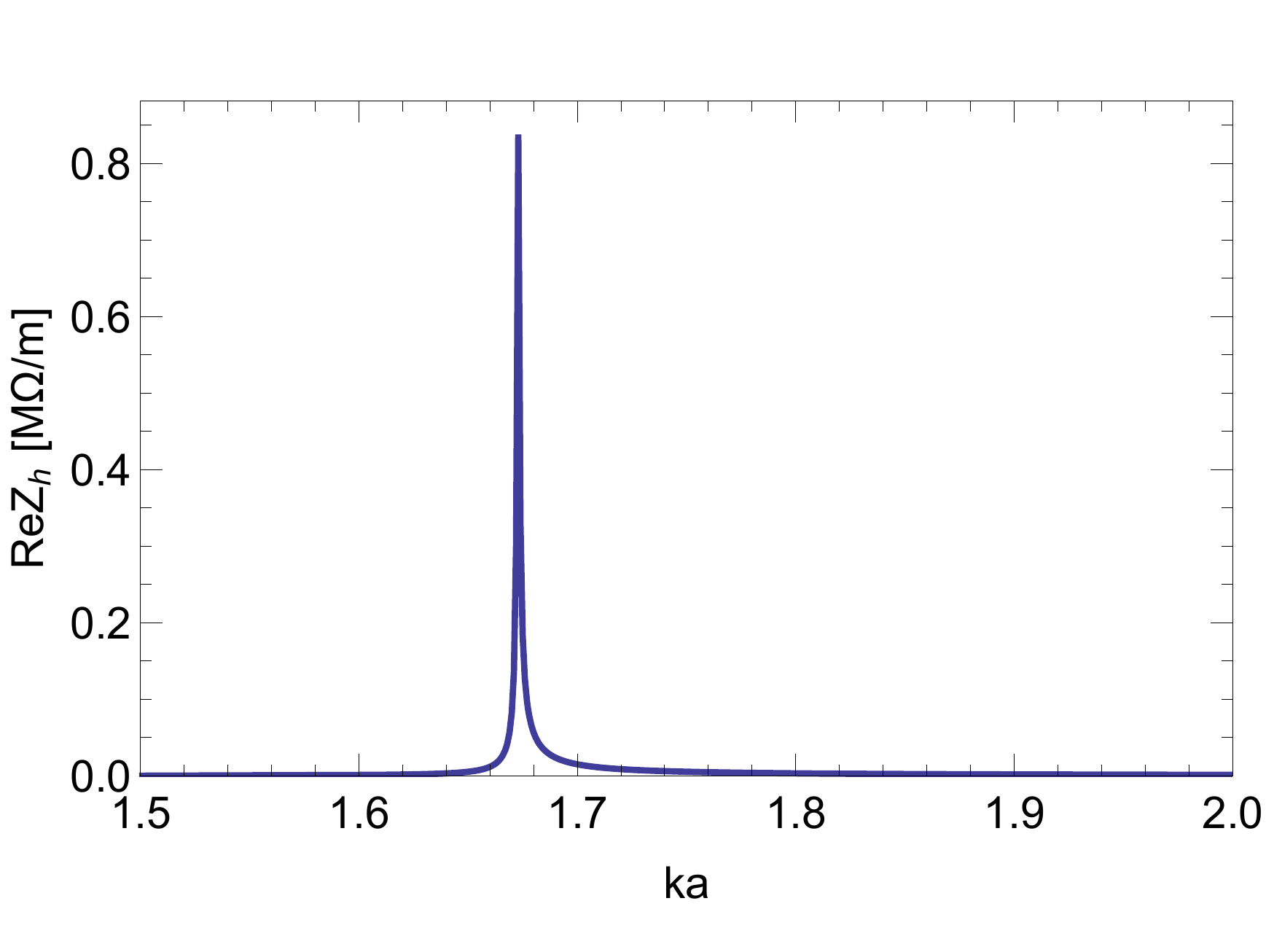}
\caption{The real part of the Joule heating impedance $Re(Z_{h})=Re(Z_{hz})+Re(Z_{hx})$ for the beam on axis in the dechirper with half-aperture $a=0.7$~mm and plate width $w=12$~mm. }
\label{Zh_w12mm_fi}
\end{figure}

Performing the numerical integrals for finite plate width $w=12$~mm, we obtain $u_h=u_{hz}+u_{hx}$ using Eqs.~\ref{losses_eq} and \ref{u2_flat_eq}. For $a=0.7$~mm, we find that the total Joule loss is a small part of the beam energy loss, $u_h=0.03u_w$ (with $u_{hz}=0.91u_h$).  Then using Eq.~\ref{ReZh_eq} we obtain the real part of the Joule heating impedances, $Z_{hz}$ and $Z_{hx}$. In Fig.~\ref{Zh_w12mm_fi} we plot the Joule impedance sum,  $Re(Z_{h})=Re(Z_{hz})+Re(Z_{hx})$. Note that on the scale of the plot, $Re(Z_{h})\approx Re(Z_{hz})$ and $Re(Z_{hx})\approx0$. As a spot check on sensitivity to conductivity, we reduced $\sigma_c$ by a factor of 4, repeated the calculation, and found that $(u_h/u_w)$ increased by 35\%.

\begin{figure}[htb]
\centering
\includegraphics[width=0.8\textwidth, trim=0mm 0mm 0mm 0mm, clip]{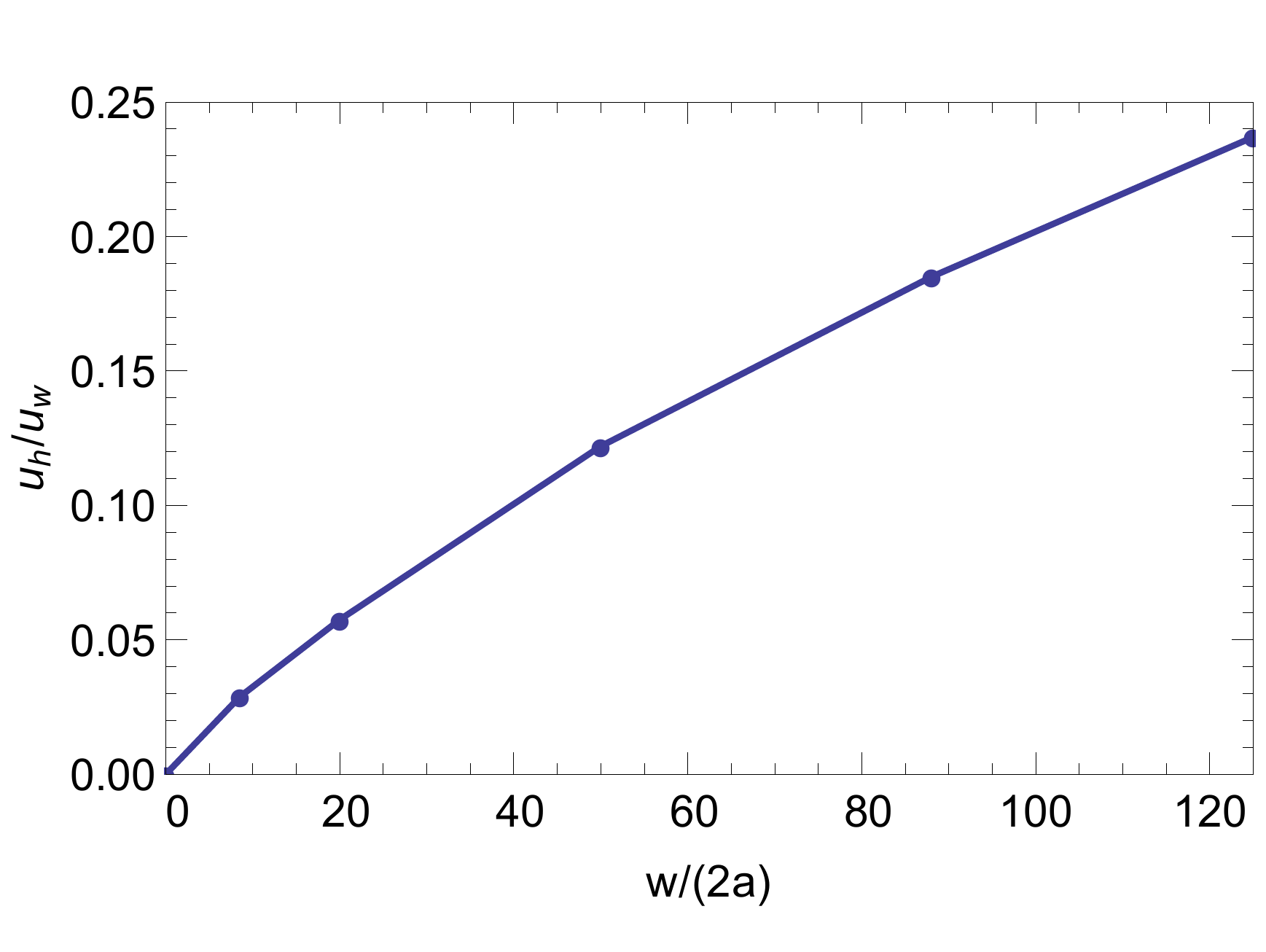}
\caption{Numerically calculated ratio of Joule loss into metal to wake loss of a point charge beam, $u_h/u_w=(u_{hz}+u_{hx})/u_w$ as function of $w/(2a)$ (plotting symbols). The wake loss for a point particle is $u_w=Q^2{\pi^2/(8a^2)}$.}
\label{p_vs_w_fi}
\end{figure}

We repeat the numerical calculation for several values of plate width $w$. In Fig.~\ref{p_vs_w_fi} we plot the Joule energy loss {\it vs.} plate width, normalized to the point charge loss of the beam, $u_w=Q^2{\pi^2/(8a^2)}$. We see that, even for $w\sim100a$, only a small part of the beam energy loss ends up as Joule heating. For our final estimate of Joule power loss of a uniform beam of full length $\ell=60$~$\mu$m we need to use the short-range, point charge wake of a beam on the axis of a flat dechirper~\cite{BaneStupakovZagorodnov16}
\begin{equation}
w_z(s)=\frac{\pi^2}{4a^2}e^{-\sqrt{s/s_{0}}}\ ,
\end{equation}
with $s_{0}=9a^2t/[8\pi \alpha(t/p)^2p^2]$ and $\alpha(x)\approx1-0.465\sqrt{x}-0.070x$. For the parameters of Table~I, the scale factor $s_0=434$~$\mu$m. For a bunch with uniform distribution of full length $\ell$, the loss factor is given by
\begin{equation}
\varkappa=\frac{1}{\ell}\int_0^\ell ds\,\left(1-\frac{s}{\ell}\right)w_z(s)\ .\label{kappa_uni_eq}
\end{equation}
Here $\varkappa=19$~kV/(pC*m), and the loss compared to a point charge beam is $\varkappa/\varkappa(0)=0.82$.

For the beam parameters of Table~I the power lost by the beam is $P_w=Q^2\varkappa f_{rep}=170$~W/m. Thus our analytical estimate of the Joule losses for the 12~mm-wide dechirper plates is the fraction $(u_h/u_w)=0.03$ of this, or $P_h=5$~W/m.

\section*{Beam Near One Plate}

There is interest in streaking the beam by inducing the transverse wakes of the dechirper, by passing the beam close to one jaw. With the beam a distance from the near wall of $b\sim0.25$~mm and from the far wall by $\gtrsim5$~mm, the second wall will no longer affect the results. The physics will be quite different than before: with two plates the impedance has a narrow resonance whose frequency depends on the plate separation $2a$; in the single plate case this parameter no longer exists. We present more details of this case, since the analysis of it is a relatively new topic.
 Note that in Ref.~\cite{one_plate_wakes} %, beginning with equations given in Ref.~\cite{analytical}, 
the wakes for a short beam passing by a single plate of a dechirper are obtained, %Note that these formulas were not given explicitly before.
and in Appendix~B of this report we use the time domain solver CST PS to test the accuracy of these results.

For the case of an ultra-relativistic beam passing by one plate of a corrugated structure, one question that comes to mind is: will the radiation generated be the same as Smith-Purcell (SP) radiation~\cite{SP}? The answer is that the radiation is not ``classical" SP radiation. With classical SP radiation, the beam energy is relatively low (a few MeVs); the radiation wavelength $\lambda=(\beta^{-1}-\cos\theta)p/n$ [$\beta$ is the ratio of the electron speed and that of light, $\theta$ is observation angle, $p$ is corrugation period, and $n$ is an integer greater than zero], {\it i.e.} the radiated wavelength is short compared to the corrugation period; and the coupling is sensitive to the detailed shape of the corrugations. For our case, in contrast, the radiation is not sensitive to $\beta$ (we can and do let $\beta=1$), the radiation period covers several corrugation periods, and the coupling is insensitive to the details of the corrugation shape.

For the Joule heating calculation we start with the equations for the magnetic fields on the wall, for a beam offset by $y$ from the axis of a two-plate dechirper (equations not shown here).  We let $y=a-b$, and then let $a\rightarrow\infty$, to obtain the fields on the walls of one plate due to a bunch passing by at distance $b$. The fields on the wall (at $y=b$) are given by 
\begin{eqnarray}
\hat H_x(q,b,\omega)&=&-\frac{4\pi Q}{c}\frac{ k |q| e^{-b |q|}}{k|q| (1+\zeta_z \zeta_x )+i \zeta_z ( q^2-k^2)}\ ,\nonumber\\
\hat H_z(q,b,\omega)&=&-i\zeta_z \hat H_x(q,b,\omega)\ .\label{Hxz_single_eq}
\end{eqnarray}
Meanwhile the electric field at the particle location (here, at $y=0$) is
\begin{equation}
\hat E_z(q,0,\omega)=-\frac{4\pi Q}{c}\frac{ \zeta_z  k |q| e^{-2 b|q|}}{k|q| (1+\zeta_z \zeta_x )+i \zeta_z ( q^2-k^2)}\ .\label{Ez_off_eq}
\end{equation}
%\begin{equation}
%\hat F_y(q,0,\omega)=\frac{4\pi Q}{c}\frac{i \zeta_z  q^2 e^{-2 b |q|}}{ k|q| (1+\zeta_z  \zeta_x )+i\zeta_z(  q^2-k^2)}
%\end{equation}

\begin{figure}[htb]
\centering
\includegraphics[width=0.8\textwidth, trim=0mm 0mm 0mm 0mm, clip]{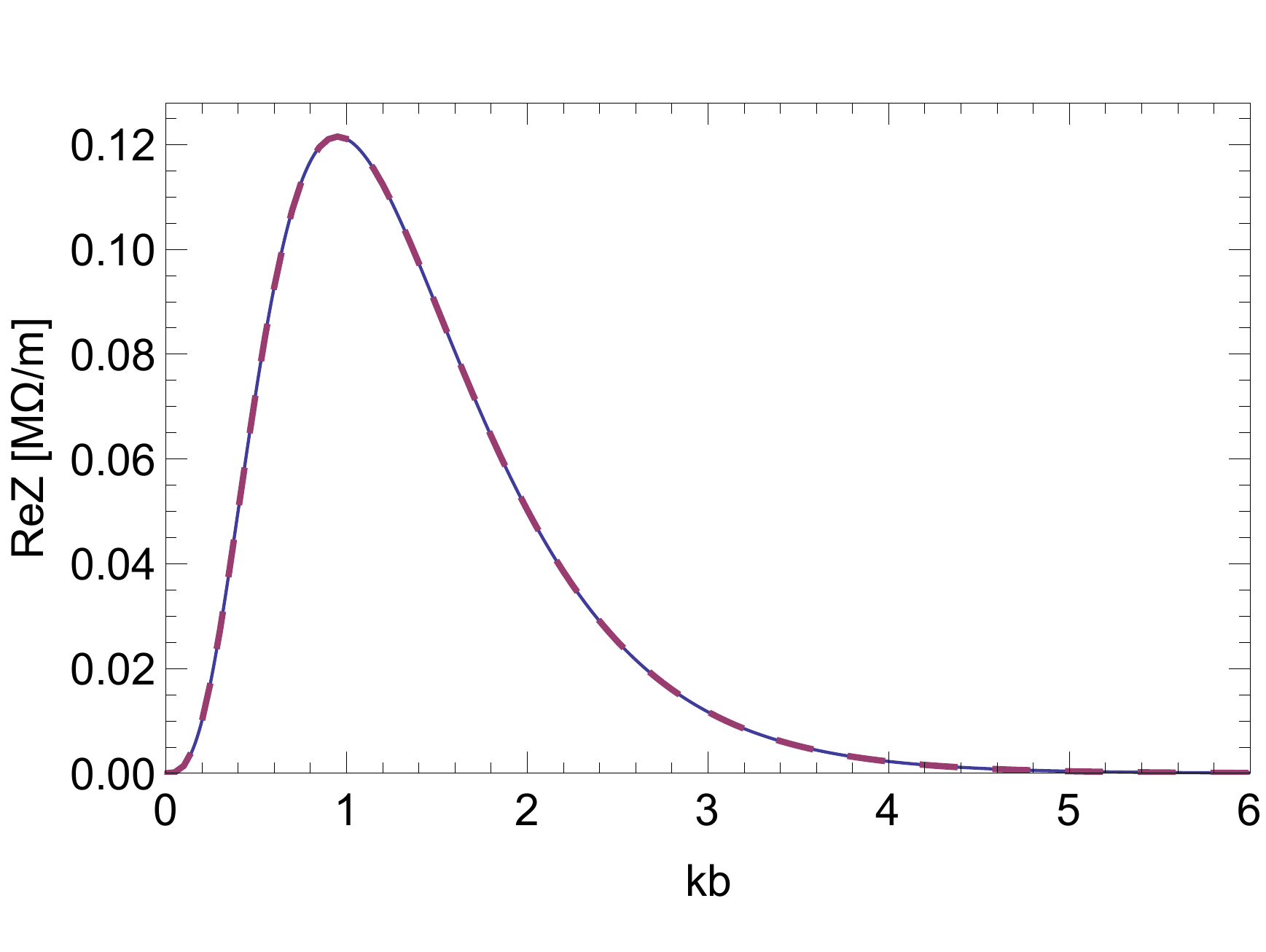}
\caption{$Re(Z)$ for beam passing by one, infinitely wide dechirper plate at a distance $b=0.25$~mm (blue). The analytical results, Eq.~\ref{Z_off_ana_eq}, is also shown (red dashes).}
\label{ReZ_off_fi}
\end{figure}

First, note that while the point charge energy loss per length of the two-plate system was $u_w=Q^2\pi^2/(8a^2)$, with $a$ the half gap, for the single plate it is only $u_w=Q^2/(2b^2)$, with $b$ the distance between the beam's path and the plate. 
In Fig.~\ref{ReZ_off_fi} we plot $Re(Z)$, where $Z=-{\tilde E_z}/Q$ (in blue). Instead of the narrow spike of the two-plate case, we now find a relatively broad peak. 

Actually, Fig.~\ref{ReZ_off_fi} can be obtained analytically. If we let the resistive terms of both $\zeta_z$ and $\zeta_x$ be zero, the integral over $q$ that needs to be performed to obtain the impedance becomes singular. However, the integral can be performed as a Cauchy integral, yielding a finite result (see Appendix~A for details). The result is
%\begin{equation}
%Re(Z)=-\frac{4\pi}{c}\frac{kqb}{1+qh}e^{-2qb}\Bigg|_{q=\frac{1}{h}\left(-1+\sqrt{1+k^2h^2}\right)}\ .\label{Z_off_ana_eq}
%\end{equation}
\begin{equation}
Re(Z)=\frac{2\pi}{c}\frac{k\xi}{1+\xi}e^{-2\xi b/h}\Bigg|_{\xi=-1+\sqrt{1+k^2h^2}}\ ,\label{Z_off_ana_eq}
\end{equation}
where $h$ is the depth of corrugation. 
Note that the frequency at the peak can be approximated as $(k)_{peak}=\sqrt{3/(2hb)}$, which is similar to the perturbation peak frequency formula for the two plate case, $(k)_{peak}=\sqrt{2/(ha)}$, with $a$ the half gap.
  The result of Eq~\ref{Z_off_ana_eq} is given by the red dashes in Fig.~\ref{ReZ_off_fi}; we see that this function is almost identical to the earlier, numerical result that included wall resistance. %Note, however, that since when $(h/b)$ becomes large the surface impedance model to the corrugated structure loses its applicability, there will be a limit size of $(h/b)$ where these results no longer apply. 

%In Fig.~\ref{ReZ_off_param_fi} we plot properties of $Re(Z)$ (Eq.~\ref{Z_off_ana_eq}) as functions of $(h/b)$: frequency at the peak $(kb)_{peak}$, full width at half maximum $(kb)_{fw}$, and peak value $Re({\hat Z})$. For small values of $(h/b)$, $(k)_{peak}=\sqrt{3/(2hb)}$ (the dashed line in the figure). This is similar to the perturbation peak frequency formula for the two plate case, $(k)_{peak}=\sqrt{2/(ha)}$, with $a$ the half gap; again the wavelength is large compared to the corrugation period. In the single plate case, for small $(h/b)$, the corresponding perturbation result for peak impedance is $Re({\hat Z})=\frac{4\pi}{c}\sqrt{\frac{27}{32}}\,e^{-3/2}/\sqrt{hb}$. 

%\begin{figure}[htb]
%\centering
%\includegraphics[width=0.8\textwidth, trim=0mm 0mm 0mm 0mm, clip]{single_plate_parameter_studyc.pdf}
%\caption{Properties of $Re(Z)$ as function of $(h/b)$ for beam passing by one dechirper wall at a distance $b$: frequency at the peak $(kb)_{peak}$, full width at half maximum $(kb)_{fw}$, and peak value $Re({\hat Z})$. The dashed line gives the small $(h/b)$ approximation, $(k)_{peak}=\sqrt{3/(2hb)}$. }
%\label{ReZ_off_param_fi}
%\end{figure}

\begin{figure}[htb]
\centering
\includegraphics[width=0.8\textwidth, trim=0mm 0mm 0mm 0mm, clip]{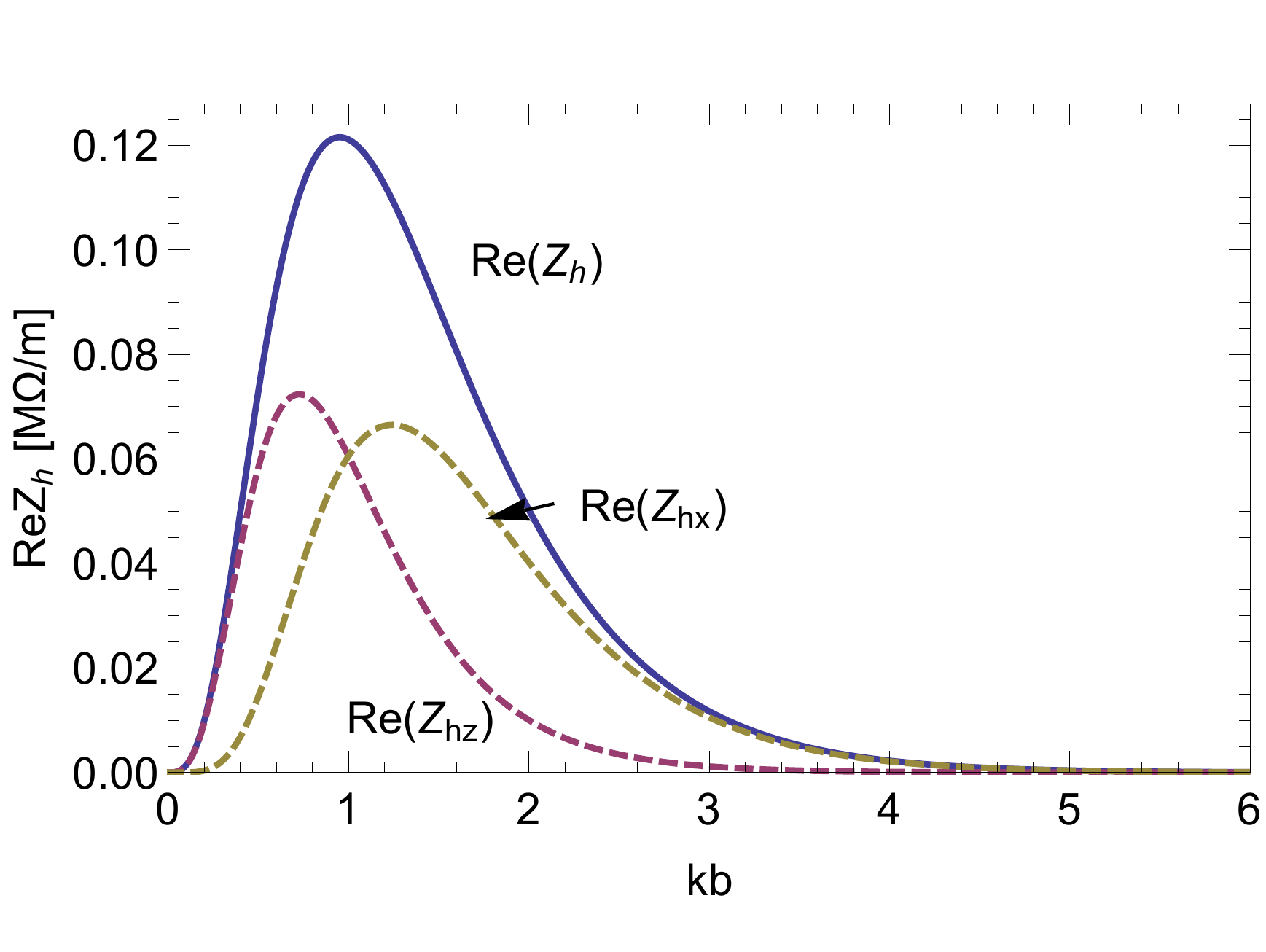}
\caption{The real part of the Joule heating impedance $Re(Z_{h})=Re(Z_{hz})+Re(Z_{hx})$ for the beam passing by a single dechirper plate, for the case of plate width $w\rightarrow\infty$ (blue curve), with the constituent parts given in dashes. The beam passes by at offset $b=0.25$~mm from the plate.  The area under $Re(Z_{hz})$ [$Re(Z_{hx})$] is 42\% [58\%] that under $Re(Z)$.}
\label{ReZh_off_fi}
\end{figure}

We next insert the single plate magnetic fields, Eq.~\ref{Hxz_single_eq}, into Eqs.~\ref{u1_winf_eq},  \ref{u2_winf_eq}, (but divided by 2, since there is only one plate) and numerically integrate to obtain the Joule heating impedance for the single, infinitely wide plate example. In Fig.~\ref{ReZh_off_fi} we show the real part of the Joule heating impedances $Re(Z_{hz})$, $Re(Z_{hx})$ (the components with wall currents aligned in, respectively, the $z$ and $x$ directions) that we obtain. Again the beam is assumed to pass by at offset $b=0.25$~mm from the plate. As expected, the total Joule heating impedance curve obtained numerically is the same, to good accuracy, as the impedance curve of Fig.~\ref{ReZ_off_fi}. The area under $Re(Z_{hz})$ [$Re(Z_{hx})$] is 42\% [58\%] that under $Re(Z)$.

\begin{figure}[htb]
\centering
\includegraphics[width=0.8\textwidth, trim=0mm 0mm 0mm 0mm, clip]{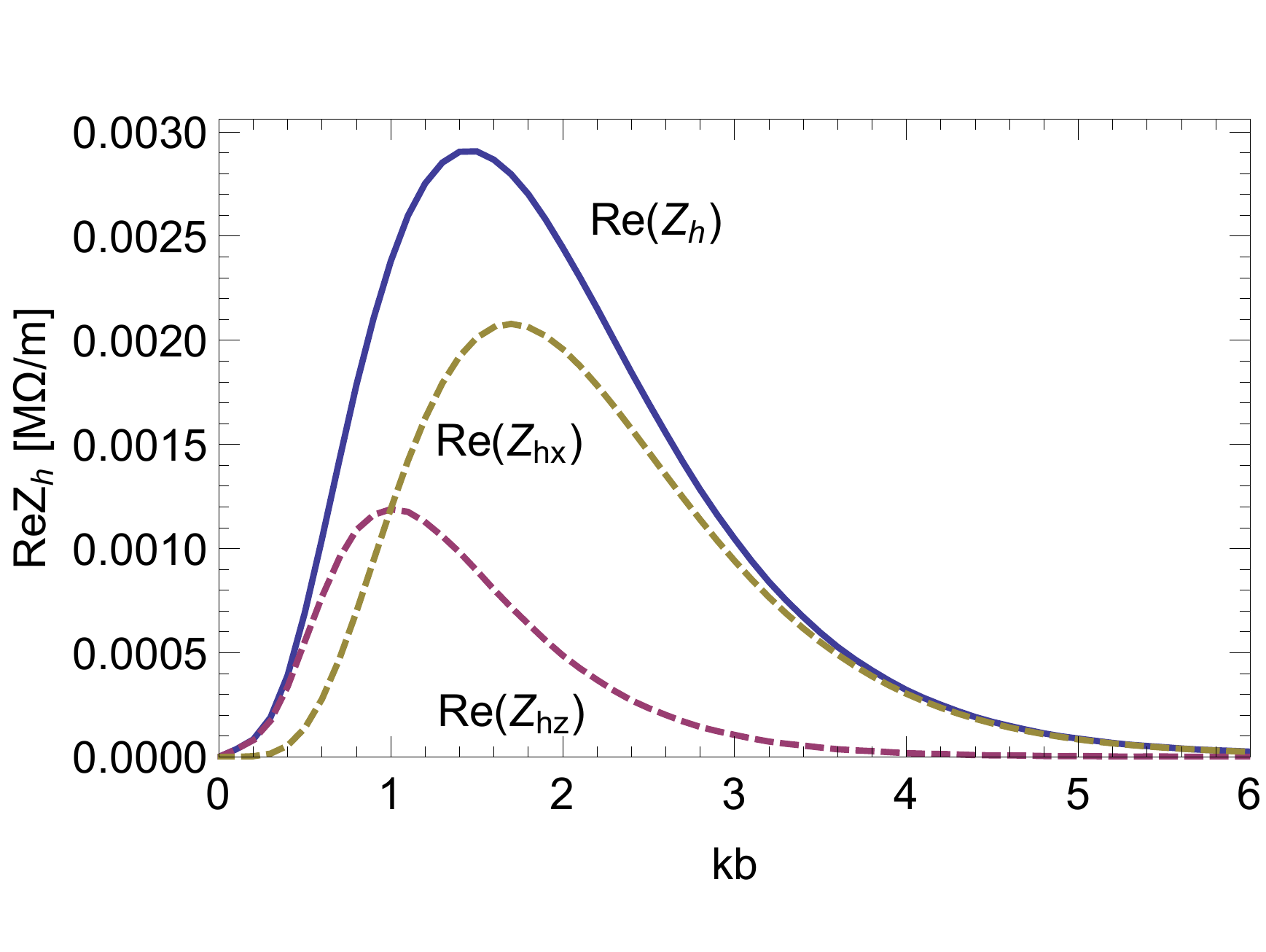}
\caption{The real part of the Joule heating impedance $Re(Z_{h})=Re(Z_{hz})+Re(Z_{hx})$ for the beam passing by a single dechirper plate, for the case of plate width $w=12$~mm (blue curve), with the constituent parts given in dashes. The beam passes by at offset $b=0.25$~mm from the plate. The area under $Re(Z_{hz})$ [$Re(Z_{hx})$] is 0.9\% [2.4\%] that under $Re(Z)$.}
\label{ReZh_off_w12mm_fi}
\end{figure}

Finally, we insert the single plate magnetic fields, Eq.~\ref{Hxz_single_eq}, into Eqs.~\ref{losses_eq},  \ref{u2_flat_eq},  (again divided by 2, since there is only one plate) and numerically integrate to find the Joule heating impedance for a beam passing by single plate with finite width $w=12$~mm.
%(at offset $b=0.25$~mm). 
The results are shown in Fig.~\ref{ReZh_off_w12mm_fi}. This time the area under the $Re(Z_{hz})$ [$Re(Z_{hx})$] curve is 0.9\% [2.4\%] of that under the impedance curve $Re(Z)$. Thus, $u_h/u_w=0.033$.

The short-range, point charge wake of a beam passing by a single plate of a flat dechirper at offset $b$ is~\cite{one_plate_wakes}
\begin{equation}
w_z(s)=\frac{1}{b^2}e^{-\sqrt{s/s_{0l}}}\ ,
\end{equation}
with $s_{0l}=2b^2t/(\pi\alpha^2p^2)$ and $\alpha=1-0.465\sqrt{t/p}-0.070(t/p)$. In Appendix~B we compare this formula with time-domain simulations, and find that the agreement is good.  For corrugation parameters of Table~I, $\alpha=0.636$. With distance from wall $b=250$~$\mu$m, $s_{0l}=98$~$\mu$m. 
For the uniform bunch distribution of Table~I, using Eq.~\ref{kappa_uni_eq}, we find that $\varkappa=48$~kV/(pC*m). The loss of a point charge beam is $\varkappa(0)=1/(2b^2)$; thus, $\varkappa/\varkappa(0)=0.67$.
The power lost by the beam is $P_w=Q^2\varkappa f_{rep}=434$~W/m. Thus our analytical estimate of the Joule losses for the 12~mm-wide dechirper plate is the fraction $(u_h/u_w)=0.033$ of this, or $P_h=14$~W/m.

\section*{Numerical Time-Domain Comparisons}

The analytical model for the short-range wakes of an LCLS-type beam passing between two jaws of the RadiaBeam/LCLS dechirper has been verified in Ref.~\cite{BaneStupakovZagorodnov16} using the finite difference, time-domain, wakefield solving program ECHO(2D)~\cite{ECHO2D}. However, in our analytical Joule heating calculations, for the case of finite-width plates, we assumed that reflections from the side walls (in $x$ for a vertical dechirper) are negligible. To see how good this approximation is, we have performed numerical, time-domain calculations using the program CST PS, both for an example with the beam on axis between two dechirper plates, and for an example with the beam passing by a single plate. For the case of two dechirper plates, the accuracy of CST PS simulations was verified by cross-checking with results using the wakefield code PBCI~\cite{PBCI}. Since the bunches are short ($z_{rms}\sim20$~$\mu$m), the catch-up distance [$z_{cu}\sim a^2/(2\sigma_z)$] is large (on the order of cm's). However, the most challenging simulation issue is the long damping time of the fields that need to be followed for the Joule loss calculation. In the double plate case this time is on the order of cm's$/c$.%, these simulations are time-consuming and challenging. %The single plate example has the extra difficulty of requiring an artificial absorbing boundary that does not reflect fields back to the jaw. 

\begin{figure}[htb]
\centering
\includegraphics[width=0.8\textwidth, trim=0mm 0mm 0mm 0mm, clip]{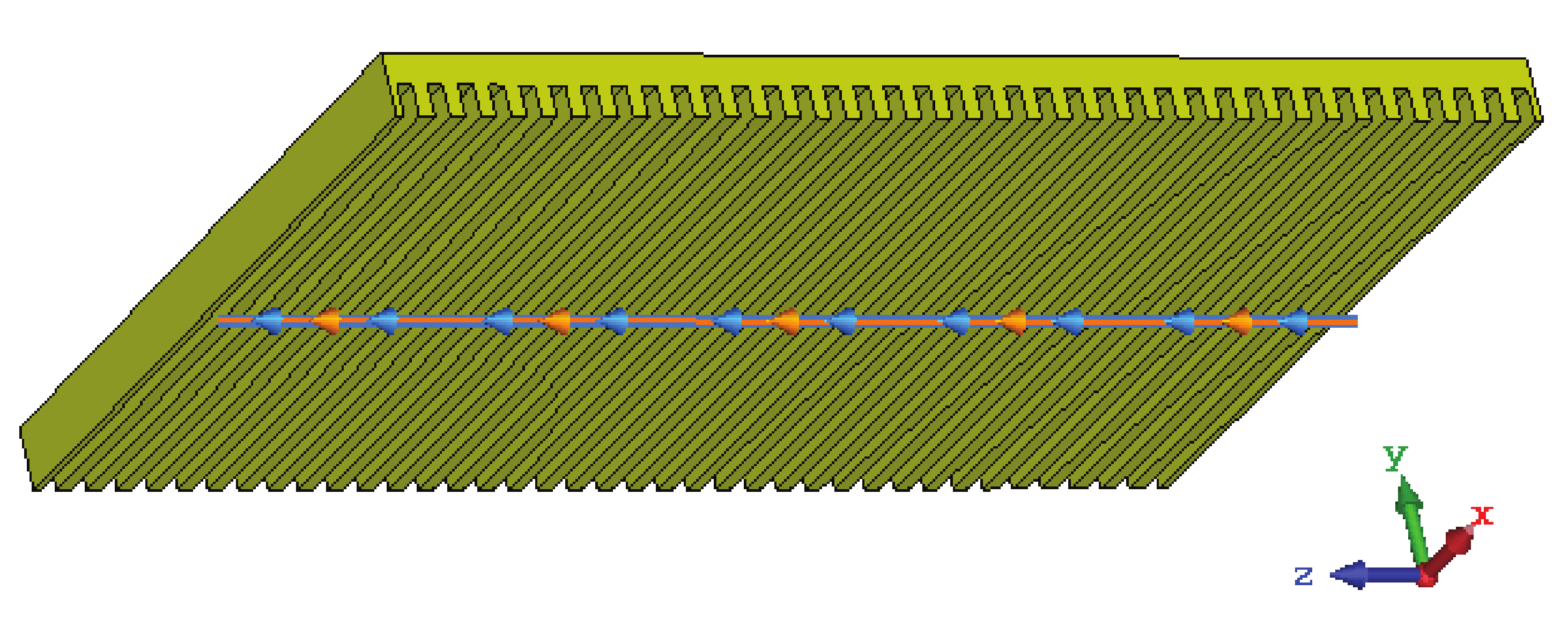}
\caption{Model used in the single-plate, numerical (CST Studio) calculations. The line and symbols indicate the beam trajectory. The plate width is $w=12$~mm, and, nominally, the beam offset is $b=0.25$~mm.}
\label{geometry_single_fi}
\end{figure}

Figure \ref{geometry_single_fi} shows the geometry of the single corrugated plate and the (nominal) beam path used in the simulations. %At the lossy surface of the plates, a surface impedace boundary condition is applied using the resistive wall impedance in (13). 
For modeling the lossy metal, a resistive wall impedance boundary condition for aluminum is applied.
The corrugated plate structure is enclosed within a larger computational box that has free-space boundary conditions applied on all sides. This is necessary for modeling field radiation and thus, for the proper computation of the Joule losses. Such boundary conditions are provided by CST PS. This is why all Joule loss results in the following were obtained using this program. To confirm that the choice of bounding box does not affect the results, we performed one calculation with changed bounding box dimensions and found that the wake obtained was left unchanged.

The simulations are performed in the time domain by tracking a single bunch along the beam path until the wake potential and Joule loss per unit length saturate to steady state. Typical plate lengths considered in the simulations are some tens of cm's.
To make the calculations manageable, for both double- and single-plate cases, we use Gaussian bunches with $\sigma_z=100$~$\mu$m, {\it i.e.} significantly longer than our nominal bunch length. The resulting mesh with the necessary numerical resolution consists of up to $1.5\times10^9$ mesh points, and it typically takes several days to complete just one simulation. Since the typical bunch frequency is much higher than the structure frequency, the ratio $(u_h/u_w)$ will be about the same for this bunch length as for the target rms bunch length (see Table~I), $\sigma_z=17$~$\mu$m.

\subsection*{Beam on Axis between Two Dechirper Plates}

\begin{figure}[htb]
\centering
\includegraphics[width=0.8\textwidth, trim=0mm 0mm 0mm 0mm, clip]{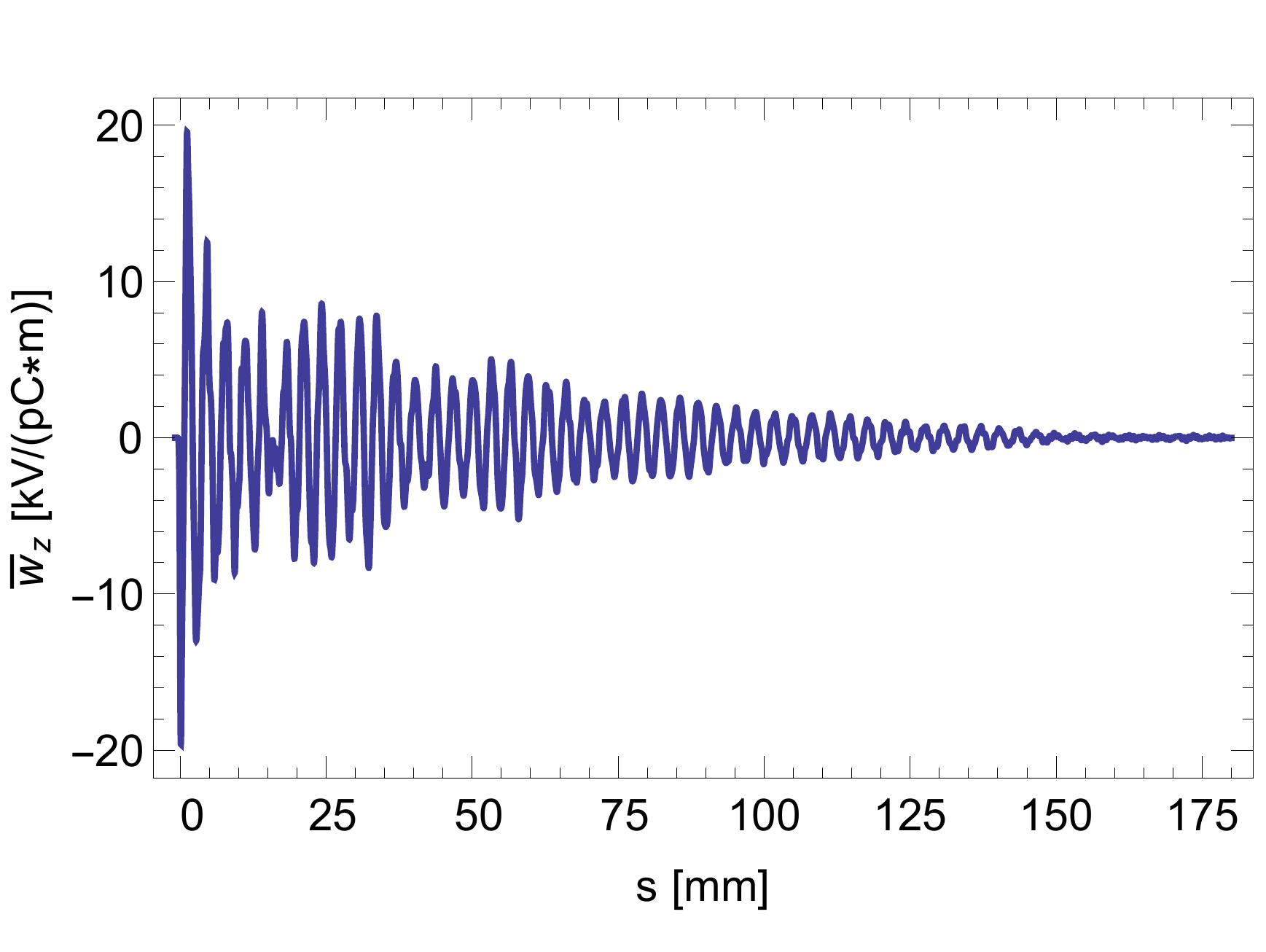}
\caption{The numerically obtained, longitudinal wake for the beam moving on-axis of a double plate dechirper. Here half-aperture $a=0.7$~mm and dechirper length $L=135$~mm; the driving bunch is Gaussian with $\sigma_z=100$~$\mu$m. Note that the wake is normalized to structure length.}
\label{Ewake_long_fi}
\end{figure}

For the two-plate case, the nominal half aperture is $a=0.7$~mm. The numerically obtained, steady-state bunch wake, for the Gaussian bunch, $\bar w_z(s)$ is given in Fig.~\ref{Ewake_long_fi}. We see that the wake damps away on a scale of $s\sim150$~mm. Fourier transforming the wake, we obtain the impedance (not shown). We find that $Re(Z)$ is given by a collection of spikes (dominated by the first one) beginning with $ka=1.36$, 1.53, 1.77; note that these values agree with results of mode matching calculations, applied to the same geometry~\cite{Zhang15}. This behavior is quite different than our analytical, perturbative solution, with its one spike at $ka=1.67$ (see Fig.~\ref{Z_winf_fi}). This mismatch was expected.

\begin{figure}[htb]
\centering
\includegraphics[width=0.8\textwidth, trim=0mm 0mm 0mm 0mm, clip]{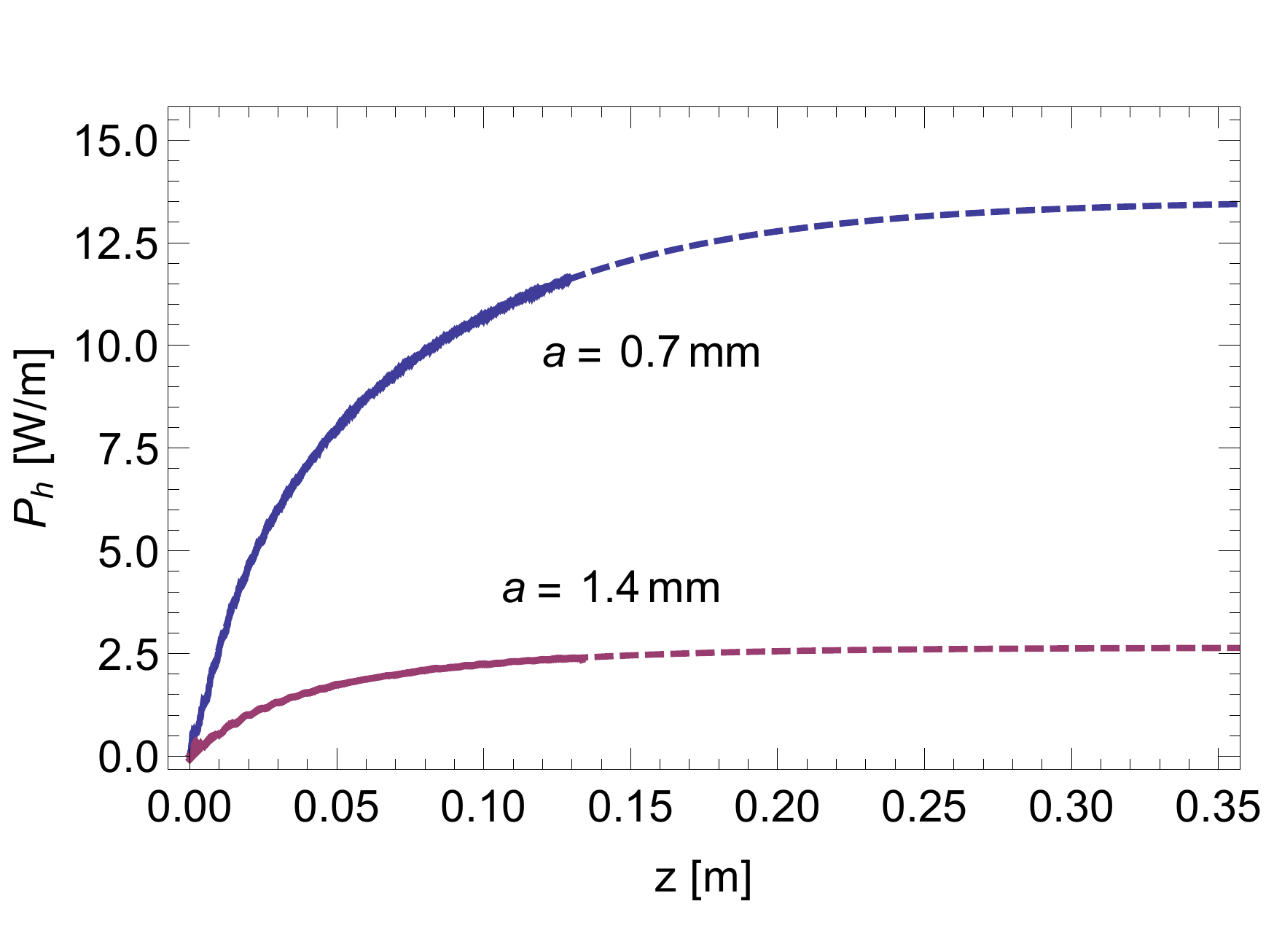}
\caption{Beam on-axis between two plates: Joule power calculations obtained by time-domain simulation, for plate half-aperture $a=0.7$~mm (blue) and $a=1.4$~mm (blue). The beam traverses the plate from its beginning, at $z=0$, to its end, at $z=0.135$~m. The dashed lines give extrapolation to steady-state. Here $Q=300$~pC, $f_{rep}=300$~kHz; the driving charge is Gaussian with $\sigma_z=100$~$\mu$m. }
\label{EJoule_fi}
\end{figure}

For the numerical Joule losses calculations, the beam passed by a two-plate dechirper of length $L=135$~mm. From the magnetic fields at the plate surfaces, the Joule heating power at each time step was obtained. Fig.~\ref{EJoule_fi} shows the results for the nominal $a=0.7$~mm case (blue, solid curve); the steady-state was obtained by extrapolation using a double exponential fitting function (the dashes). The final result is $P_h=13.5$~W/m. The wake power loss for the $\sigma_z=100$~$\mu$m Gaussian bunch $P_w=108$~W/m. Thus, the ratio of Joule heating and wake energy, according to the numerical calculation, is $(u_h/u_w)_{\rm num}=P_h/P_w=0.125$, which is a factor of 4 larger than the analytical result obtained above, $(u_h/u_w)_{\rm ana}=0.030$. Our best estimate for the Joule power loss for the LCLS-II beam, ${\bar P}_h$, is obtained taking $(u_h/u_w)_{\rm num}$ and multiplying it with the wake power (analytically obtained above) for the short, uniform LCLS-II  bunch shape, $(P_w)_{ana}$; {\it i.e.}
\begin{equation}
{\bar P}_h=(u_h/u_w)_{\rm num}(P_w)_{\rm ana}\ .
\end{equation}
For the nominal case we obtain ${\bar P}_h=(0.125)(170\ {\rm W/m})=21.0$~W/m. 

As a final word on the two-plate calculation, note that a second case was also simulated, with $a=1.4$~mm.  In this case, $(h/a)=0.35$ is half the size of before, and we find that the impedance (not shown) is closer to the analytical one. The simulation finds that the Joule power $P_h=2.64$~W/m (see Fig.~\ref{EJoule_fi}, the red curve). The energy ratio $(u_h/u_w)_{\rm num}=0.065$, which is a factor of 2.5 larger than the analytical value $(u_h/u_w)_{ana}=0.025$, rather than the factor of 4 we had before. For $a=1.4$~mm, $(P_w)_{\rm ana}=46$~W/m, and our best estimate of Joule power loss becomes ${\bar P}_h=(0.065)(46\ {\rm W/m})=
3.0$~W/m.

\subsection*{Beam Passing by a Single Corrugated Plate}
 
\begin{figure}[htb]
\centering
\includegraphics[width=0.8\textwidth, trim=0mm 0mm 0mm 0mm, clip]{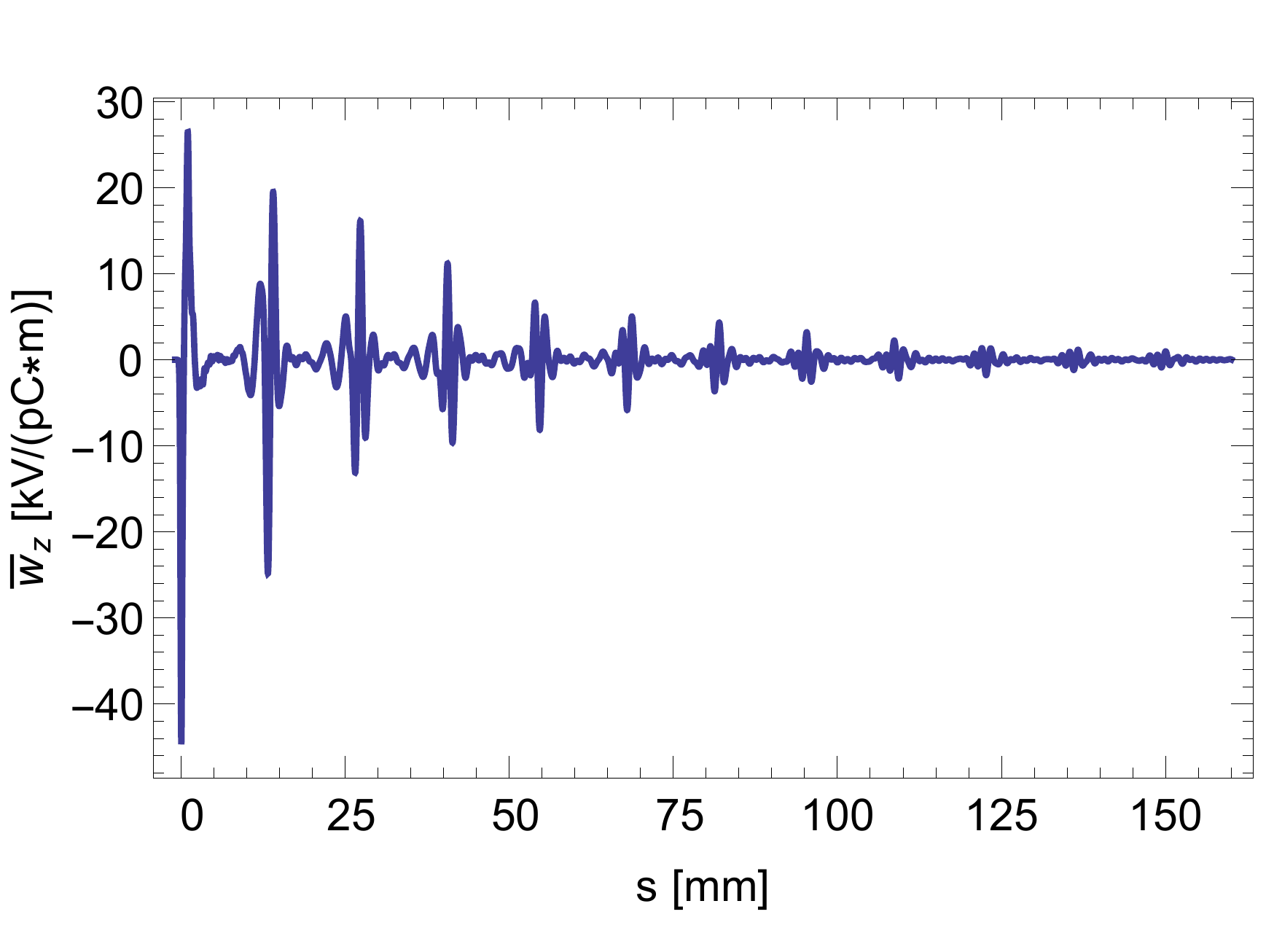}
\caption{The numerically obtained, longitudinal wake for a beam moving past a single-plate dechirper. Here the beam offset $b=0.25$~mm and dechirper length $L=115$~mm; the driving bunch is Gaussian with $\sigma_z=100$~$\mu$m. Note that the wake is normalized to structure length. }
\label{Eswake_long_fi}
\end{figure}

The numerically obtained wake for a 100~$\mu$m Gaussian bunch passing at distance $b=0.25$~mm from the plate is shown in Fig.~\ref{Eswake_long_fi}. Here we see that the wake dies out after $s\sim120$~mm and that there are reflections from the sides of the plate. Note that since the period in the reflections is $\sim13.5$~mm (and not equal to the plate width, $w=12$~mm), we infer that the group velocity of the waves moving sideways is $v_g=(12/13.5)c=0.89c$. 

\begin{figure}[htb]
\centering
\includegraphics[width=0.8\textwidth, trim=0mm 0mm 0mm 0mm, clip]{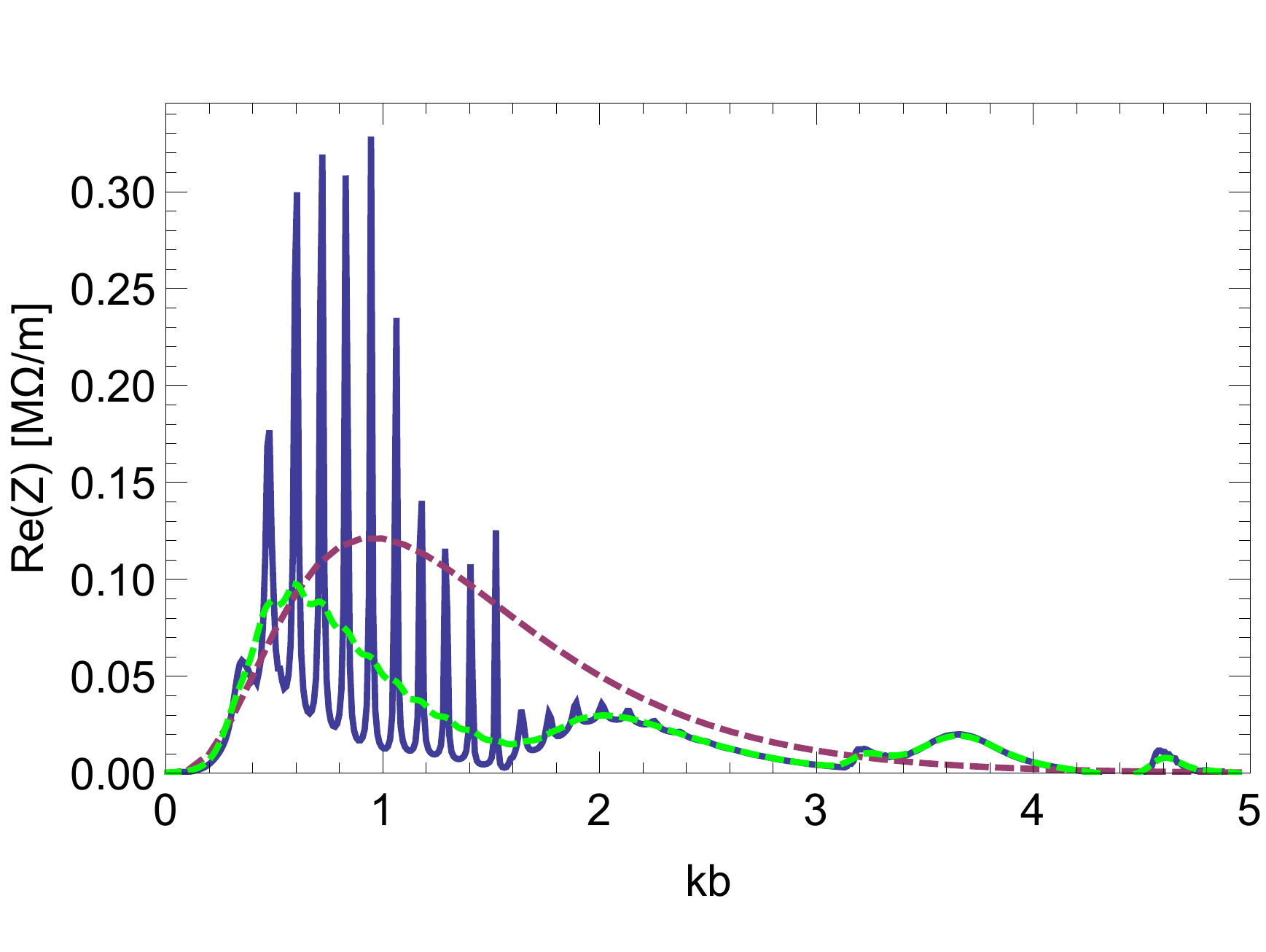}
\caption{The real value of the impedance ${\rm Re}(Z)$ for the single plate example (blue curve).
% To obtain this, the longitudinal wake (shown in Fig.~\ref{Eswake_long_fi}) was Fourier transformed and multiplied by $e^{k^2\sigma_z^2/2}$. 
The broad-band impedance obtained from the same wake is given by green dashes. The analytical perturbation result (Fig.~\ref{ReZ_off_fi}), is given by red dashes. }
\label{EsZ_fi}
\end{figure}

The real part of the impedance ${\rm Re}(Z)$ for the single plate example is given in Fig.~\ref{EsZ_fi} (blue curve).  To obtain this, the longitudinal wake (Fig.~\ref{Eswake_long_fi}) was Fourier transformed and multiplied by $e^{k^2\sigma_z^2/2}$. The narrow, evenly-spaced spikes in the impedance [with spacing $\Delta f\approx c/(13.5\ {\rm cm})$] are due to the reflections in the wake. A short bunch, however, cannot resolve these spikes. To generate a ``broad-band impedance", one that is easier to compare with our analytical result, we multiplied the wake by a Gaussian form factor, with rms length $\sigma=5$~mm, before Fourier transforming. The resulting impedance is given by green dashes in the figure. Our analytical result (Fig.~\ref{ReZ_off_fi}) is given in red dashes. Although the low and higher frequency behavior of the red and green curves agree well, the numerical peak is narrower and the frequency of the peak is lower than the analytical one. This disagreement appears to be a consequence of the corrugation parameters not being in the perturbative regime: here $(h/b)=2$, which is not small compared to 1.

\begin{figure}[htb]
\centering
\includegraphics[width=0.8\textwidth, trim=0mm 0mm 0mm 0mm, clip]{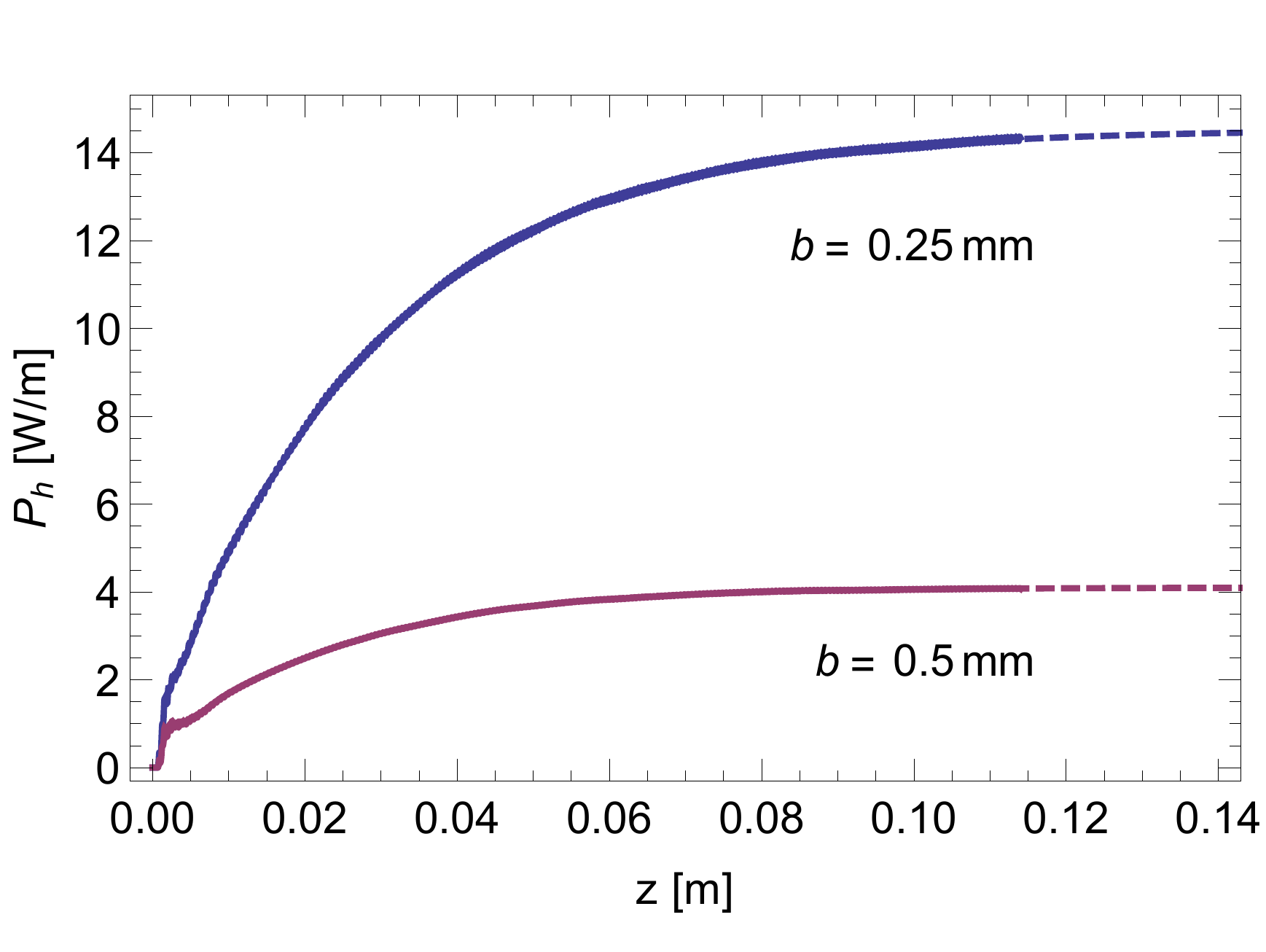}
\caption{Single plate: numerically obtained, Joule power calculations for beam offsets $b=0.25$~mm (blue) and  $b=0.5$~mm (red). The beam traverses the plate from its beginning, at $z=0$, to its end, at $z=0.115$~m. The dashed lines give extrapolation to steady-state. Here $Q=300$~pC, $f_{rep}=300$~kHz; the driving charge is Gaussian with $\sigma_z=100$~$\mu$m.}
\label{EJoules_fi}
\end{figure}

For Joule loss simulations, the beam was passed by a single-plate dechirper of length $L=115$~mm. Fig.~\ref{EJoules_fi} shows the results for the nominal $b=0.25$~mm case (blue, solid curve), and the extrapolation to steady-state (the dashes). The steady-state result is $P_h=14.5$~W/m. The wake power loss for the $\sigma_z=100$~$\mu$m Gaussian bunch $P_w=264$~W/m. Thus, the ratio of Joule heating and wake energy according to the numerical calculation is $(u_h/u_w)_{\rm num}=P_h/P_w=0.055$; this is a factor of 1.67 larger than the analytical result obtained above, $(u_h/u_w)_{\rm ana}=0.033$. This is due to the multiple reflections of the wakefield from the sides of the plate that are not considered in the analytical model. Our best estimate of the Joule power loss for the LCLS-II beam, ${\bar P}_h=24.0$~W/m.

More single plate simulations were performed for larger beam offsets: $b=0.5$, 1.0, 1.5~mm. As we move to ever smaller values of $(h/b)$, the numerically obtained (broad-band) impedance agrees better with the analytical one. The numerical energy ratio $(u_h/u_w)$, however, remains about a factor of 2 larger than the analytical one.
For $b=0.5$~mm, the simulations find that the Joule power $P_h=4.0$~W/m (see Fig.~\ref{EJoule_fi}, the red curves). The energy ratio $(u_h/u_w)_{\rm num}=0.035$, and our best estimate of Joule power loss becomes ${\bar P}_h=%(0.065)(46\ {\rm W/m})=
4.6$~W/m.
Finally, note that our nominal, two-plate and single-plate results are summarized in Table~II.

\begin{table}[hbt]
   \centering
   \caption{Summary of Joule heating calculations for the LCLS-II dechirper, giving case; wake power lost by beam, $(P_w)_{\rm ana}$; ratio of energy in Joule heating and beam energy loss, analytical calculation,  $(u_h/u_w)_{\rm ana}$; Joule power loss, analytical calculation, $(P_h)_{\rm ana}$; energy ratio, according to numerical calculation, $(u_h/u_w)_{\rm num}$; and our best estimate of Joule losses, ${\bar P}_h=(u_h/u_w)_{\rm num}(P_w)_{ana}$. Both cases assume the high charge scenario, with $Q=300$~pC and $f_{rep}=100$~kHz; the bunch shape is taken as uniform, with total length $\ell=60$~$\mu$m.  }
   \begin{tabular}{||l|c|c|c|c|c||}\hline\hline 
       Case &  $(P_w)_{\rm ana}$[W/m]  &  $(u_h/u_w)_{\rm ana}$ &  $(P_h)_{\rm ana}$[W/m]  &  $(u_h/u_w)_{\rm num}$ &  ${\bar P}_h$ [W/m] \\ \hline\hline 
Two plates, $a=0.7$~mm &170 &0.030   & 5  &  0.125  &21.0  \\ \hline
Single plate, $b=0.25$~mm & 435  &  0.033 & 14  & 0.055  & 24.0 \\
         \hline \hline 
   \end{tabular}
   \label{table2_tab}
\end{table}

\section*{Conclusions}

We have performed Joule power loss calculations for the RadiaBeam/LCLS-II dechirper, whose engineering details---for example concerning the cooling required---are still being finalized. We have investigated the configurations of the beam on-axis between the two plates, for chirp control, and for the beam especially close to one plate, for use as a fast kicker. Our calculations involve an analytical model that uses a surface impedance approach, valid for perturbatively small dechirper parameters
%, but here applied to cases where the parameters were not small. 
In addition, our model ignores effects of field reflections at the sides of the dechirper plates, and is thus expected to underestimate the Joule losses. The analytical results were also tested by numerical, time-domain simulations using computer programs in CST Studio and PBCI. We find that most of the wake power lost by the beam is radiated out to the sides of the plates. While our theory can be applied to the LCLS-II dechirper with large gaps,
for the nominal apertures we are not in the perturbative regime and the reflection contribution to Joule losses is not negligible. 
With input from computer simulations, we estimate the Joule power loss (assuming bunch charge of $300$~pC, repetition rate of $100$~kHz) is 21~W/m for the case of two plates, and 24~W/m for the case of a single plate.

The single-plate configuration of a dechirper has, until now, received little attention in the literature. 
In this report we have presented also the impedance of a beam passing by a single corrugated plate. Also, in Appendix~B we have numerically confirmed that the analytical expressions for short-range wakes found in Ref.~\cite{one_plate_wakes} are valid when $(h/b)\ll1$, where $h$ is depth of corrugation and $b$ is distance of beam from plate.  In fact, we showed that the longitudinal and dipole (but not quadrupole) wakes agree well even for $(h/b)\lesssim2$. 

\section*{Acknowledgments}

We thank I. Zagorodnov for helpful discussions on the subject of the short-range wakefields in the dechirper. Work supported by the U.S. Department of Energy, Office of Science, Office of Basic Energy Sciences, under Contract No. DE-AC02-76SF00515.

%appendix
%\appendix\newpage\markboth{Appendix}{Appendix}
\renewcommand{\thesection}{\Alph{section}}
\numberwithin{equation}{section}

\begin{appendices}

%Appendix A

\section{Derivation of $Re(Z)$ for beam passing by a single dechir\-per plate}

\begin{figure}[htb]
\centering
\includegraphics[width=0.65\textwidth, trim=0mm 0mm 0mm 0mm, clip]{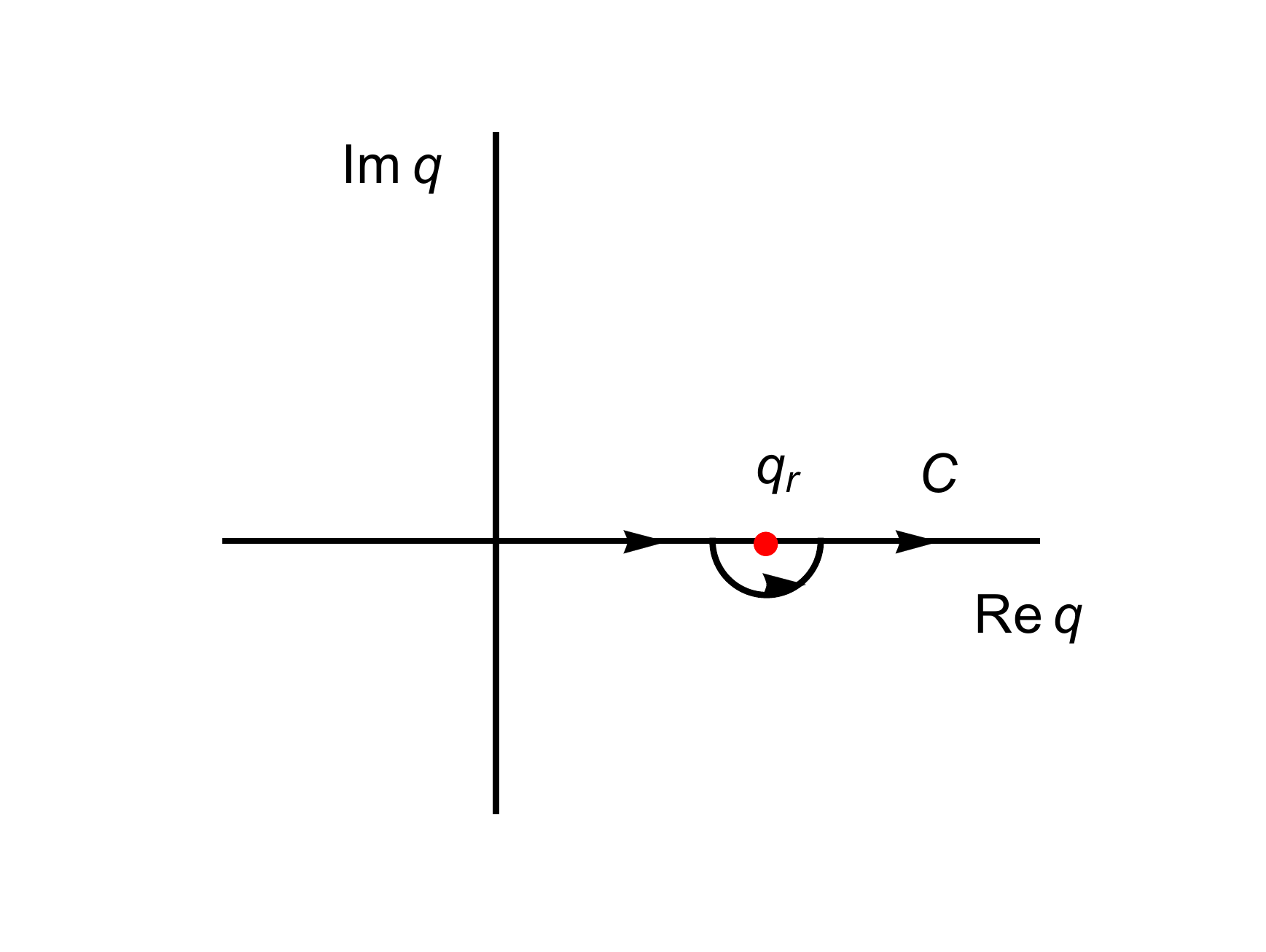}
\caption{Integration path $C$ in the complex plane of variable $q$.}
\label{fig:15}
\end{figure}

The impedance is
    \begin{align}
    Z(\omega)
    =
    -
    \frac{1}{2\pi Q}
    \int_{-\infty}^\infty
    dq 
    \hat E_z(q,0,\omega)\ ,
    \end{align}
where (see Eq.~\ref{Ez_off_eq}) 
\begin{equation}
\hat E_z(q,0,\omega)=-\frac{4\pi Q}{c}\frac{ \zeta_z  k |q| e^{-2 b|q|}}{k|q| (1+\zeta_z \zeta_x )+i \zeta_z ( q^2-k^2)}\ .
\end{equation}
We need to calculate the following integral
    \begin{align}
    Z(\omega)
    =
    \frac{4}{c}
    \zeta_z  k
    \int_{0}^\infty
    dq \,
    \frac{  q e^{-2 bq}}
    {kq (1+\zeta_z \zeta_x )+i \zeta_z ( q^2-k^2)}\ 
    \end{align}
 (where we have used the fact that the integrand is symmetric with respect to $q$).
In the limit when the resistive term $\zeta_{rw}\to 0$, we have $\zeta_x=0$ and
    $
    \zeta_z
    =
    - \frac{1}{2}ihk
    $
and the integral reduces to
    \begin{align}\label{eq:a4}
    Z(\omega)
    =
    -
    \frac{4i}{c}
    hk^2
    \int_{0}^\infty
    dq \,
    \frac{ q e^{-2 bq}}
    {2kq+hk ( q^2-k^2)}\ .
    \end{align}
It is easy to see that the denominator of the integrand vanishes at
\begin{equation}
q=q_r=\frac{-1+\sqrt{1+h^2k^2}}{h}\ ,
\end{equation}
and the integrand has a pole at this point. This pole has to be bypassed in the complex plane (of variable $q$) or, equivalently, the integration path needs to be shifted from the real axis. The direction of the shift can be found by analyzing the position of the pole when $\zeta_{rw}$ is small, but not equal to zero. This analysis shows that for non-zero $\zeta_{rw}$ the pole has a positive imaginary part, which means that in the limit $\zeta_{rw}\to 0$ the integration path should be modified as shown in Fig.~\ref{fig:15}.

We are interested in calculating the real part of the impedance, ${\rm Re}(Z)$. Because of the imaginary factor in front of the integral~\eqref{eq:a4}, the real part of $Z$ is equal to the half-residue of the integrand at $q=q_r$. A straightforward calculation of the residue yields
\begin{equation}
\mathrm{Re}(Z)=\frac{2\pi}{c}\frac{k\xi}{1+\xi}e^{-2\xi b/h}\Bigg|_{\xi=-1+\sqrt{1+k^2h^2}}\,.
\label{Z_off_anab_eq}
\end{equation}
 
%Appendix B
\section{Confirmation of Short-Range Wake Formulas for the Case of a Beam Passing by a Single Corrugated Plate}

The short-range wake formulas for a beam passing between the two corrugated plates of a dechirper were verified by time-domain, finite difference simulations using the computer program ECHO(2D)~\cite{BaneStupakovZagorodnov16}. Then the short-range wake formulas for a beam passing by a single corrugated plate were derived in Ref.~\cite{one_plate_wakes} without such verification. Here we present time-domain, finite difference simulations using the time-domain Maxwell Equation solver of CST Studio with the goal of testing the single-plate, short-range wake formulas.

The analytical, longitudinal point-charge wake for a beam passing by a single corrugated plate at distance $b$ is~\cite{one_plate_wakes}
\begin{equation}
w_z(s)=\frac{1}{b^2}e^{-\sqrt{s/s_{0l}}}\ ,
\end{equation}
with $s_{0l}=2b^2t/(\pi\alpha^2p^2)$ and $\alpha=1-0.465\sqrt{t/p}-0.070(t/p)$. For corrugation parameters of Table~I, $\alpha=0.636$. With distance from wall $b=250$~$\mu$m, $s_{0l}=98$~$\mu$m. 
The dipole wake (we assume the plate is above the beam at vertical offset $b$)
\begin{equation}
w_{yd}(s)=\frac{2}{b^3}s_{0y}\left[1-\left(1+\sqrt{\frac{s}{s_{0y}}}\right)e^{-\sqrt{s/s_{0y}}}\right]\ ,
\end{equation}
with $s_{0y}=8b^2t/(9\pi\alpha^2p^2)$. Here  $s_{0y}=44$~$\mu$m.
The quad short-range wake is given by $w_{yq}(s)=\frac{3}{2b}w_{yd}(s)$.
The wake of a Gaussian bunch of length $\sigma_z$ can be obtained from the point-charge wake $w(s)$ by convolution; {\it e.g.}
\begin{equation}
{\bar w}_z(s)=-\frac{1}{\sqrt{2\pi}\,\sigma_z}\int_0^\infty ds' w_z(s')\exp\left[{-\frac{1}{2}\frac{({s-s'})^2}{{\sigma_z^2}}}\right]\ .
\end{equation}
The Gaussian bunch wakes ${\bar w}_{yd}(s)$ and  ${\bar w}_{yq}(s)$ are obtained in the analogous manner, except without the overall minus sign.

\begin{figure}[htb]
\centering
\includegraphics[width=0.55\textwidth, trim=0mm 0mm 0mm 0mm, clip]{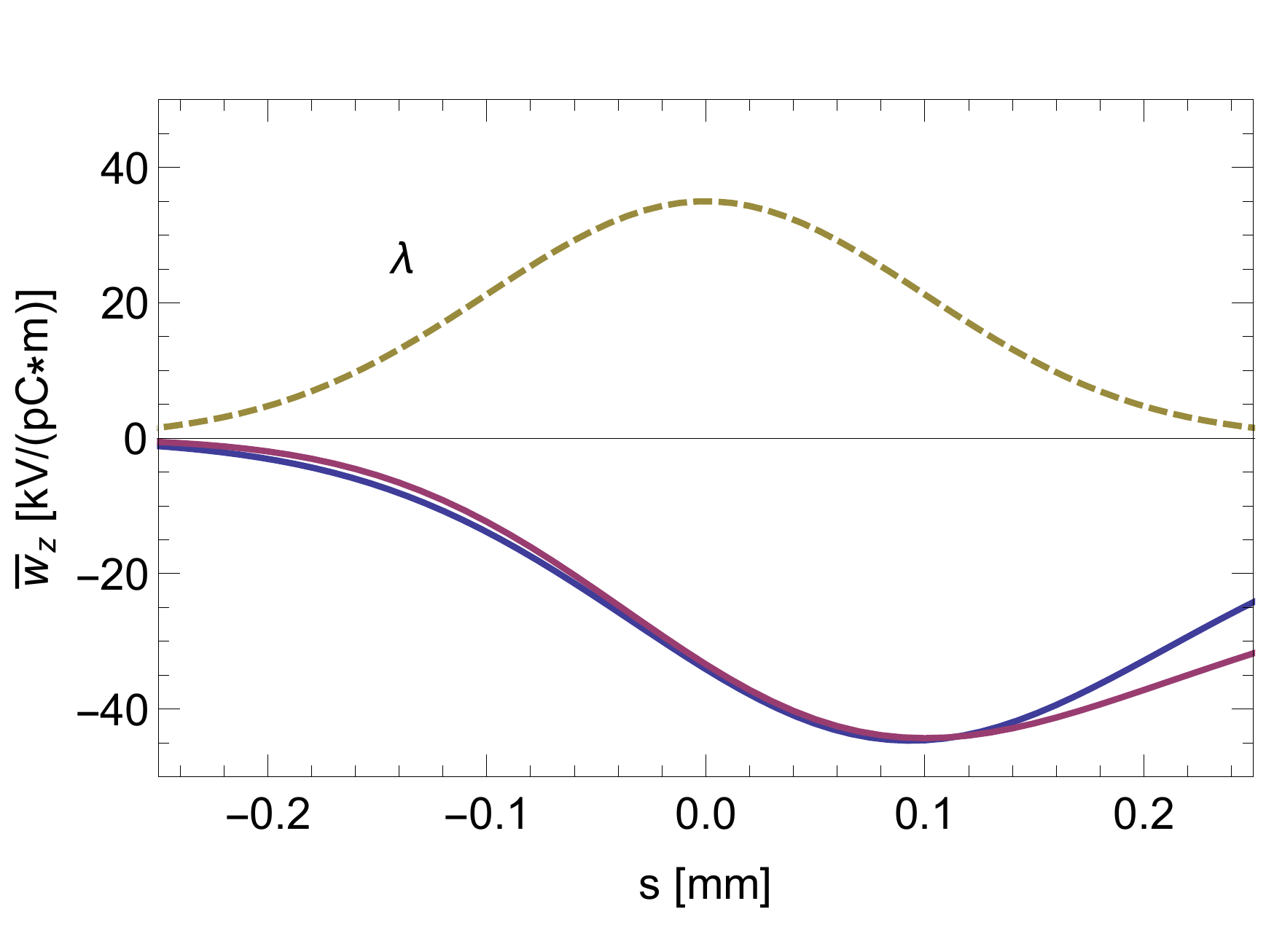}
\includegraphics[width=0.55\textwidth, trim=0mm 0mm 0mm 0mm, clip]{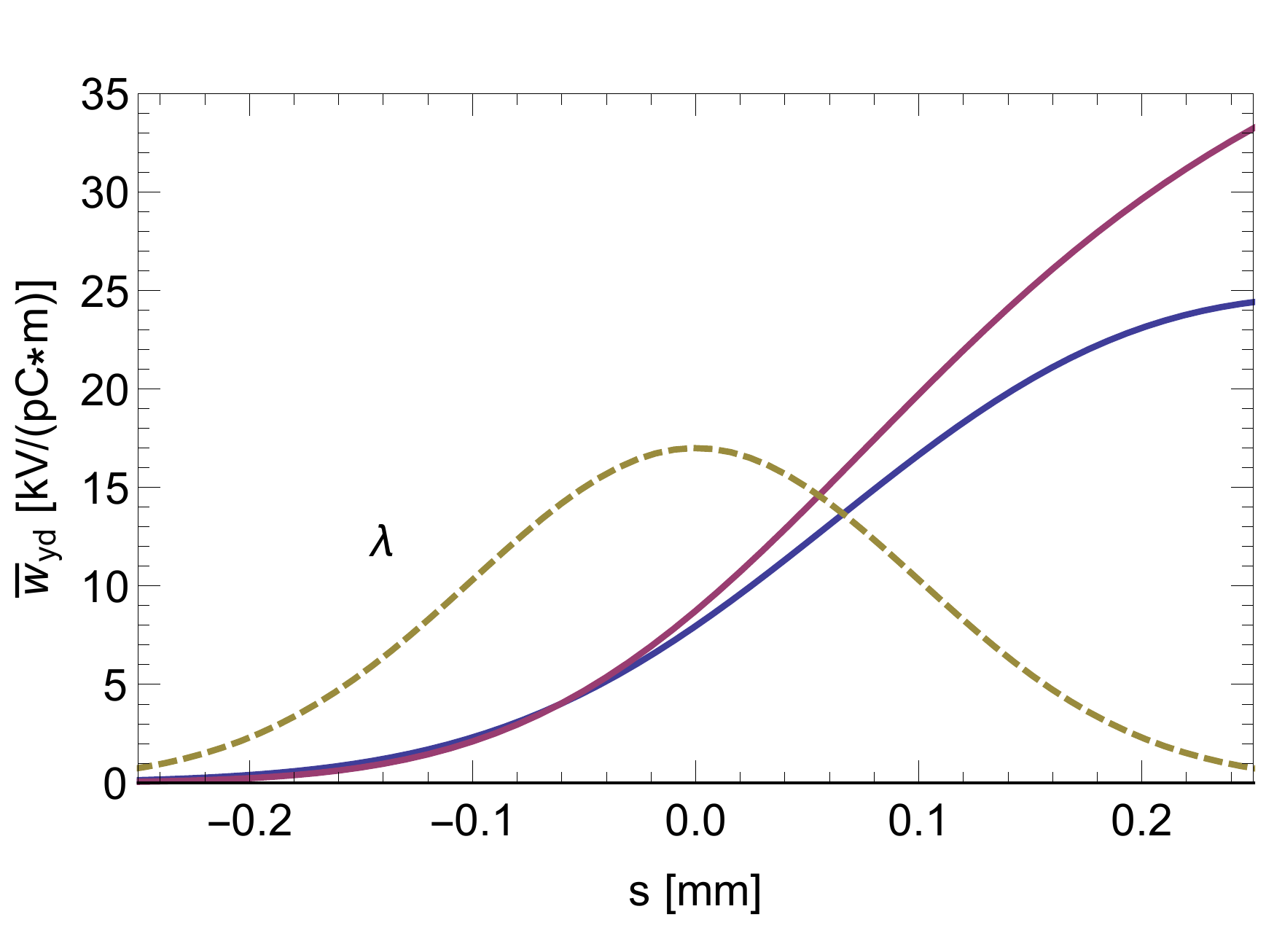}
\includegraphics[width=0.55\textwidth, trim=0mm 0mm 0mm 0mm, clip]{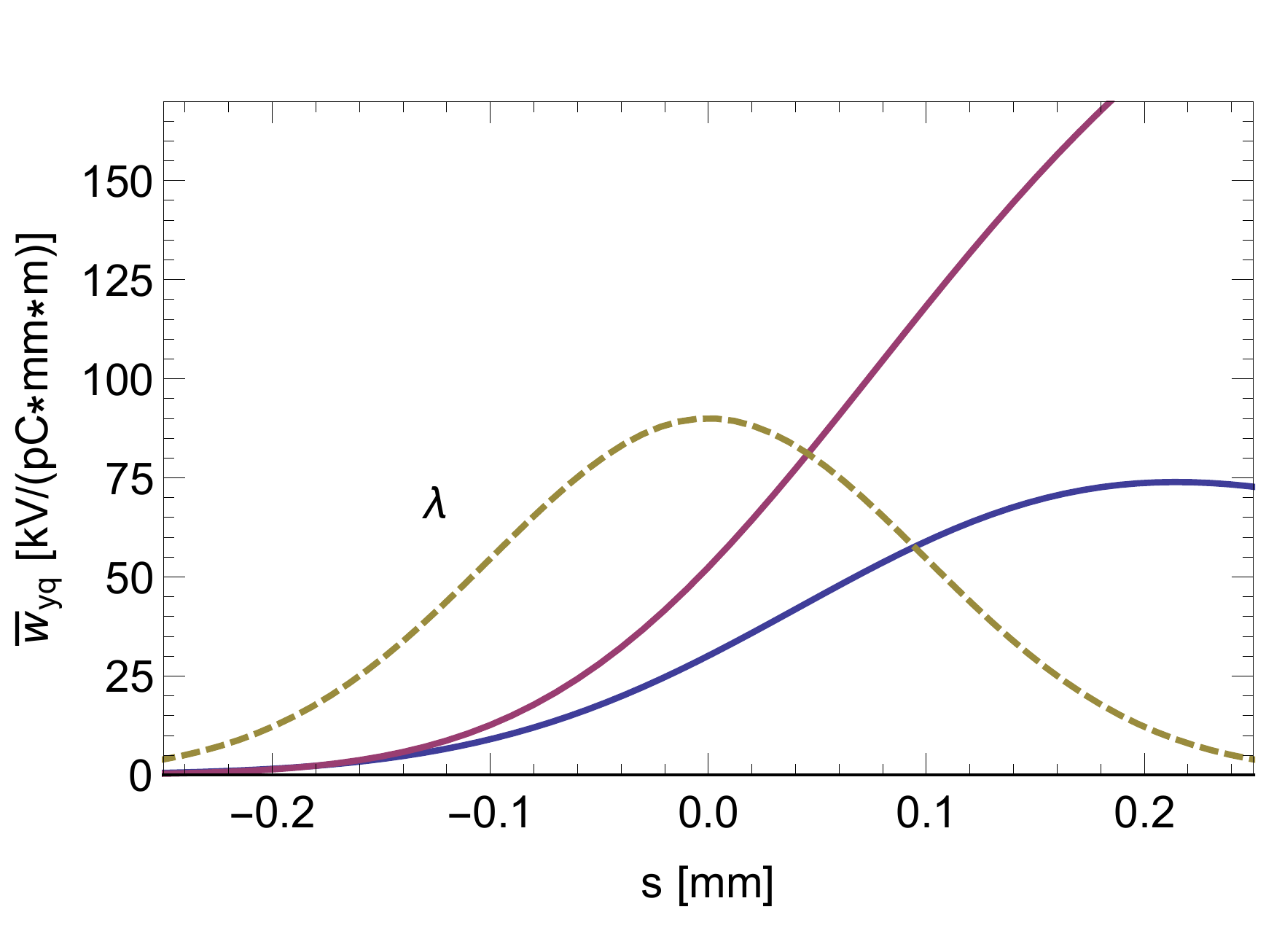}
\caption{Short-range wakes for a Gaussian bunch passing by a single plate of the dechirper at offset $b=250$~$\mu$m: ${\bar w}_z$ (top), ${\bar w}_{yd}$ (middle), and ${\bar w}_{yq}$ (bottom). The CST results are given in blue, those of the analytical model in red. The plate width used in the simulations is $w=12$~mm, the bunch length $\sigma_z=100$~$\mu$m. The shape of the bunch distribution $\lambda(s)$, with the head to the left, is also shown.}
\label{wakes_comp_fi}
\end{figure}

In Fig.~\ref{wakes_comp_fi} we plot the longitudinal, dipole, and quad wakes of a Gaussian bunch (with $\sigma_z=100$~$\mu$m) passing by a single dechirper plate at distance $b=250$~$\mu$m, as obtained numerically (in blue) and compare with the analytical results (in red). The bunch distribution is also shown, with the bunch head to the left. We see that the longitudinal and dipole bunch wakes agree well over the bunch; the quad analytical wake, however, is significantly larger than the numerical result. 

\begin{figure}[htb]
\centering
\includegraphics[width=0.55\textwidth, trim=0mm 0mm 0mm 0mm, clip]{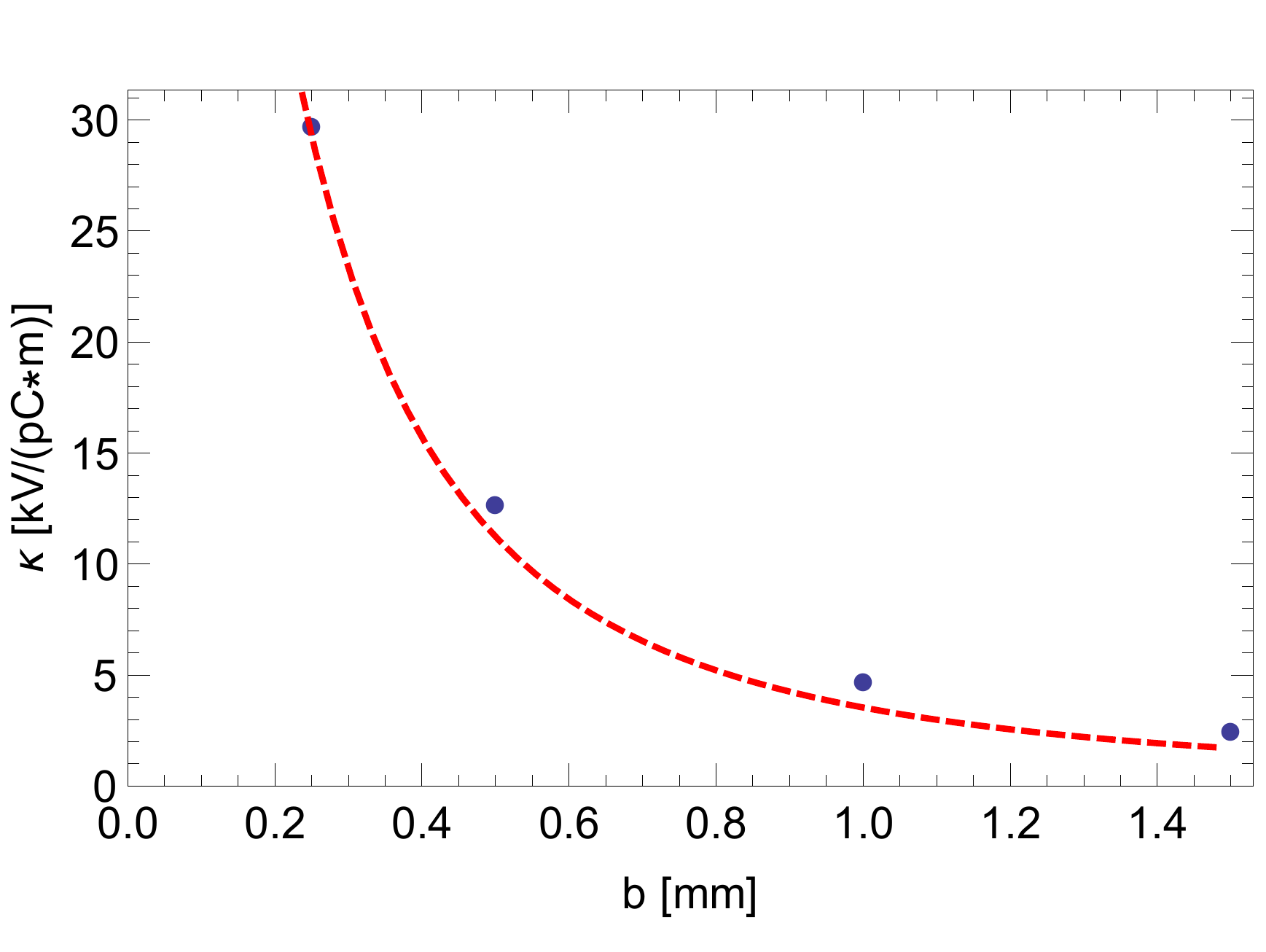}
\includegraphics[width=0.55\textwidth, trim=0mm 0mm 0mm 0mm, clip]{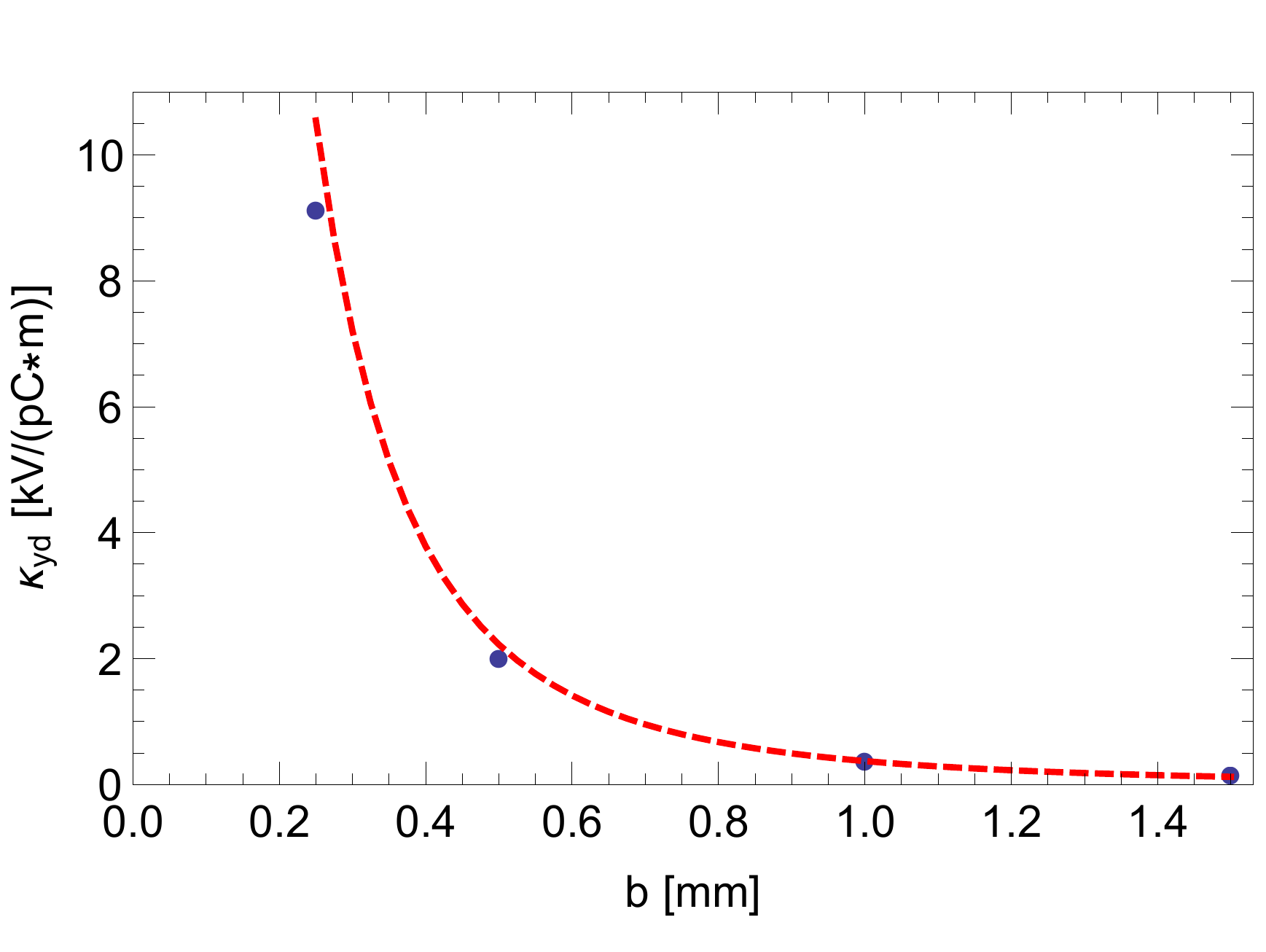}
\includegraphics[width=0.55\textwidth, trim=0mm 0mm 0mm 0mm, clip]{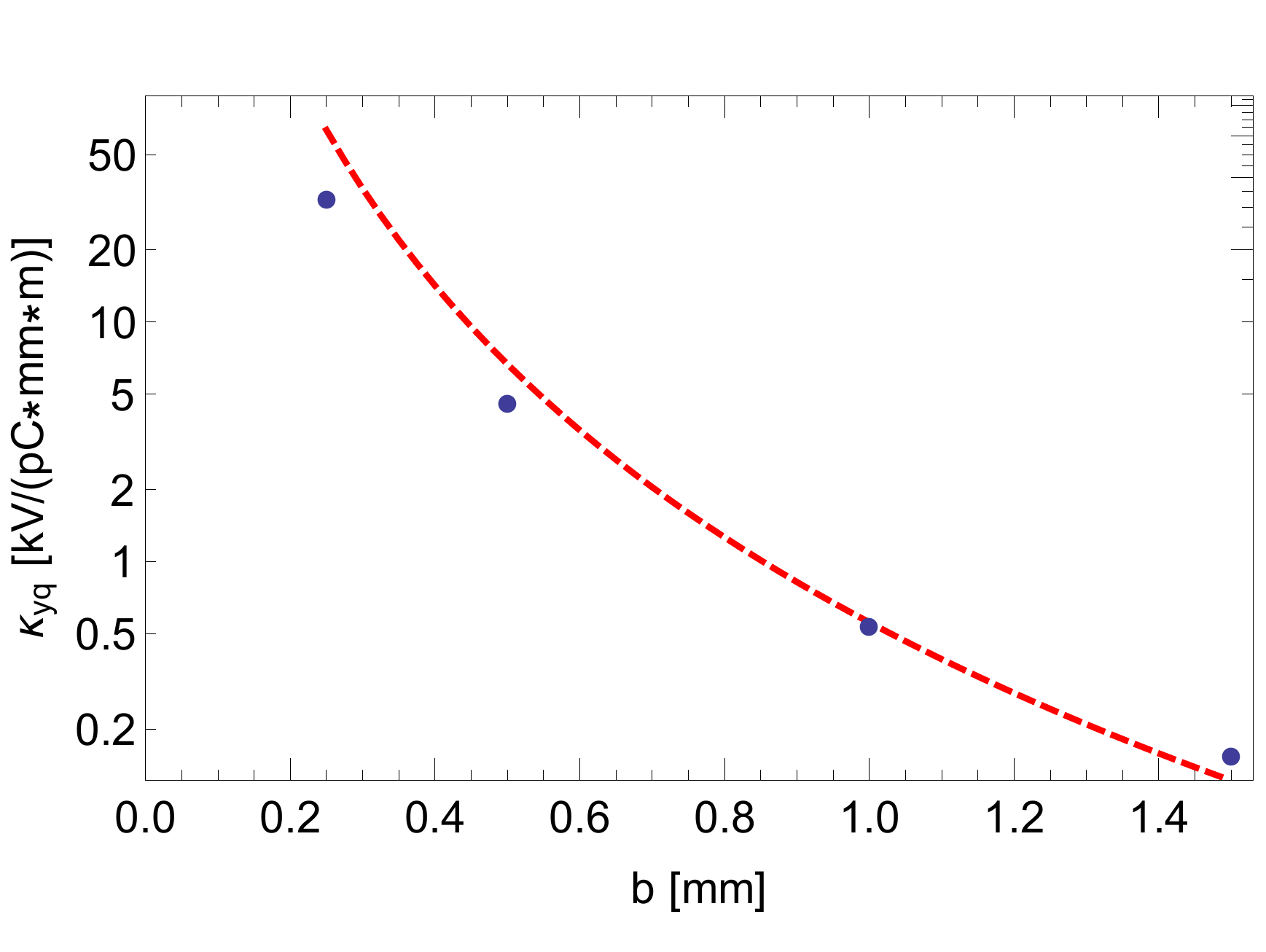}
\caption{Single plate loss factor $\varkappa$ (top), dipole kick factor $\varkappa_{yd}$ (middle), and quad kick factor $\varkappa_{yq}$ (bottom) as functions of distance of the beam from the wall $b$, showing the CST results (blue symbols) and those of the analytical model (red dashes). The bunch is Gaussian with length $\sigma_z=100$~$\mu$m; the plate width in the simulations is $w=12$~mm. Note that the ordinate of the last plot is given on a log scale.}
\label{kappa_comp_fi}
\end{figure}

The numerical simulations were repeated for bunch offsets $b=0.5$, 1, 1.5~mm.
Averaging the bunch wake over the Gaussian bunch distribution we obtain the loss factor or kick factors. 
In Fig.~\ref{kappa_comp_fi} we compare the loss and kick factors for the four beam offsets (blue symbols) with the analytical result (red, dashed curves). (The quad kick factors are shown on a log scale, to better see the comparison at the larger values of $b$.) We see that $\varkappa$ and $\varkappa_{yd}$ agree well for all offsets. However, $\varkappa_{yq}$, at $b=250$ (500) $\mu$m has an analytical solution that is 90\% (45\%) larger than the numerical one; at the larger two offsets, however, the agreement is again quite good. 
%We should not be surprised: at $b=250$ (500)~$\mu$m, $(h/b)=2$ (1), which is not small compared to 1; we are clearly not in the perturbative regime. The fact that $\varkappa$ and $\varkappa_{yd}$ agree well at the lower two values of $b$ we believe is somewhat accidental. However, note that for the larger values of $b$, where $(h/b)=0.5$, 0.33, we are closer to the perturbative regime, and the average values of the three wakes agree well with the numerical results. 

In conclusion, we have verified that the short-range longitudinal, dipole, and quad analytical formulas for wakes of a single plate agree well with the results of numerical simulations, provided we are in the perturbative regime, {\it i.e.} $(h/b)\ll1$. Furthermore, the longitudinal and dipole wakes agree well even for $(h/b)\lesssim2$.

\end{appendices}


\begin{thebibliography}{1}

\bibitem{Bane12}
K. Bane and G. Stupakov, ``Corrugated Pipe as a Beam Dechirper," Nucl. Instrum. Meth. {\bf A}690 (2012) 106--110.

\bibitem{Guetg16}
M. Guetg, et al, ``Commissioning of the RadiaBeam/SLAC Dechirper," Proc. IPAC2016, Busan, Korea, 2016, p. 809.

\bibitem{Lutman16}
A. Lutman, et al, ``Fresh-slice multicolour X-ray free-electron lasers," Nature Photonics, {\bf 10} (2016) 745--750.

\bibitem{BaneStupakov12b}
K. Bane and G. Stupakov, ``Surface impedance formalism for a metallic beam pipe with small corrugations," Phys. Rev. ST Accel. Beams {\bf 15}, 124401 (2012).

\bibitem{Zeta}
K. Bane and G. Stupakov, ``Using surface impedance for calculating wakefields in flat geometry," Phys. Rev. ST Accel. Beams {\bf 18}, 034401 (2015).

\bibitem{Erratum}
K. Bane and G. Stupakov, ``Erratum: Using surface impedance for calculating wakefields in flat geometry," Phys. Rev. Accel. Beams, {\bf 19}, 039901 (2016).

\bibitem{BaneStupakov03}
K. Bane and G. Stupakov, ``Impedance of a rectangular beam tube with small corrugations," Phys. Rev. ST Accel. Beams {\bf 6}, 024401 (2003).

\bibitem{Zhang15}
Z. Zhang, et al,  ``Electro beam energy chirp control with a rectangular corrugated structure at the Linac Coherent Light Source," Phys. Rev. ST Accel. Beams {\bf 18}, 010702 (2015).

\bibitem{Novokhatski15}
A. Novokhatski, ``Wakefield potentials of corrugated structures," Phys. Rev. ST Accel. Beams {\bf 18}, 104402 (2015).

%\bibitem{BaneStupakov16} 
%K. Bane and G. Stupakov, ``Dechirper wakefields for short bunches," Nucl. Instrum. Meth. A820 (2016) 156--163.

\bibitem{BaneStupakovZagorodnov16}
K. Bane, G. Stupakov, I. Zagorodnov, ``Analytical formulas for short bunch wakes in a flat dechirper,"  Phys. Rev. Accel. Beams, {\bf 19}, 084401 (2016).

\bibitem{one_plate_wakes}
K. Bane, G. Stupakov, I. Zagorodnov, ``Wakefields of a beam near a single plate in a flat dechirper," SLAC-PUB-16881, November 2016.

\bibitem{CST}
CST-Computer Simulation Technology, CST Particle Studio. {http://www.cst.com} 

\bibitem{PBCI}
E. Gjonaj, et al, New Journal of Physics, {\bf 8}, 285 (2006).

\bibitem{SP}
S.J. Smith and E.M. Purcell, ``Visible Light from Localized Surface Charges Moving across a Grating," Phys. Rev. {\bf 92}, 1069 (1953).

\bibitem{ECHO2D}
I. Zagorodnov, K. Bane, G. Stupakov, ``Calculation of wakefields in 2D rectangular structures," Phys. Rev. ST Accel. Beams {\bf 18}, 104401 (2015).

\bibitem{AChao}
A. Chao, {\it Physics of Collective Beam Instabilities in High Energy Accelerators}, (J. Wiley \& Sons, New York, 1993), Chap. 2.


%\bibitem{Bane12}
%K. Bane and G. Stupakov, ``Corrugated pipe as a beam dechirper," Nucl. Instrum. Meth. A690 (2012) 106--110.

%\bibitem{THz}
%K. Bane and G. Stupakov, ``Terahertz radiation from a pipe with small corrugations," Nucl. Instrum. Meth. A677 (2012) 67--73.

%\bibitem{BaneStupakov03}
%K. Bane and G. Stupakov, ``Impedance of a rectangular beam tube with small corrugations," Phys. Rev. ST Accel. Beams {\bf 6}, 024401 (2003).


%\bibitem{analytical}
%K. Bane, G. Stupakov, I. Zagorodnov, ``Analytical formulas for short bunch wakes in a flat dechirper," Phys Rev Accel and Beams {\bf 19}, 084401 (2016).


\end{thebibliography}
\end{document}